\documentclass[10pt,a4paper]{article}
\pdfoutput=1
\usepackage{graphicx, xspace}
\graphicspath{{plots/}}
\usepackage[dvipsnames]{xcolor}
\usepackage[utf8]{inputenc}
\usepackage[T1]{fontenc}
\usepackage{amsmath}
\usepackage{amssymb}
\usepackage{booktabs}
\usepackage{multirow}
\usepackage{enumitem}
\usepackage{listings}
\usepackage{caption}
\usepackage{setspace}
\usepackage[numbers,sort&compress]{natbib}
\usepackage{hyperref}
\usepackage{titlesec}
\usepackage[a4paper,left=1.in,right=1.in,top=1.in,bottom=1.in]{geometry}

\usepackage[most]{tcolorbox}
\tcbuselibrary{listings, breakable}

\definecolor{promptbg}{RGB}{230,240,255}
\definecolor{answerbg}{RGB}{240,255,240}
\definecolor{promptborder}{RGB}{100,150,220}
\definecolor{answerborder}{RGB}{80,180,100}
\definecolor{elapsedbg}{RGB}{255,243,215}
\definecolor{elapsedborder}{RGB}{200,140,50}

\newtcolorbox{promptbox}[1][]{
  colback=promptbg, colframe=promptborder,
  boxrule=1.2pt,
  fonttitle=\bfseries\footnotesize,
  fontupper={\small\setstretch{1}},
  title=User Prompt,
  left=5pt, right=5pt, top=3pt, bottom=3pt,
  breakable, #1
}

\newtcolorbox{answerbox}[1][]{
  colback=answerbg, colframe=answerborder,
  boxrule=1.2pt,
  fonttitle=\bfseries\footnotesize,
  fontupper={\small\setstretch{1}},
  title=Final Answer,
  left=5pt, right=5pt, top=3pt, bottom=3pt,
  breakable, #1
}
\newtcolorbox{elapsedbox}[1][]{
  colback=elapsedbg, colframe=elapsedborder,
  boxrule=1.2pt,
  fontupper={\small\setstretch{1}},
  left=5pt, right=5pt, top=3pt, bottom=3pt,
  #1
}
\newcommand{\testprompt}[1]{\begin{promptbox}#1\end{promptbox}}
\newcommand{\testanswer}[1]{\begin{answerbox}#1\end{answerbox}}
\newcommand{\testelapsed}[1]{\begin{elapsedbox}\textbf{Elapsed time:} #1\end{elapsedbox}}

\usepackage{tikz}
\usepackage{tikz-feynman}

\tikzfeynmanset{compat=1.1.0}
\usetikzlibrary{arrows.meta, positioning}

\hypersetup{
  colorlinks=true,
  linkcolor=Maroon,
  urlcolor=Maroon,
  citecolor=Maroon
}

\newcommand{\gev}{\ensuremath{\text{GeV}}\xspace}
\newcommand{\TeV}{\ensuremath{\text{TeV}}\xspace}

\newcommand{\MG}{\textsc{MadGraph5\_aMC@NLO}\xspace}
\newcommand{\Pythia}{\textsc{Pythia8}\xspace}
\newcommand{\Root}{\textsc{Root}\xspace}
\newcommand{\pyroot}{\textsc{PyRoot}\xspace}
\newcommand{\lhe}{\textsc{LHE}\xspace}
\newcommand{\event}{\textsc{EventLoop}\xspace}
\newcommand{\lhereader}{\textsc{LHEReader}\xspace}
\newcommand{\rooagent}{\textsc{RooAgent}\xspace}
\newcommand{\FastJet}{\textsc{FastJet}\xspace}
\newcommand{\MadAnalysis}{\textsc{MadAnalysis5}\xspace}
\newcommand{\langraph}{\textsc{LangGraph}\xspace}
\newcommand{\langchain}{\textsc{LangChain}\xspace}
\newcommand{\pypi}{\textsc{PyPI}\xspace}
\newcommand{\Claude}{\textsc{Claude}\xspace}
\newcommand{\OpenAI}{\textsc{OpenAI}\xspace}
\newcommand{\Ollama}{\textsc{Ollama}\xspace}
\newcommand{\GPTfourone}{\texttt{GPT-4.1}\xspace}
\newcommand{\DeepSeekVthree}{\texttt{DeepSeek-V3}\xspace}
\newcommand{\ClaudeSonnet}{\texttt{Sonnet~4.6}\xspace}

\makeatletter\def\l@subsubsection#1#2{}\makeatother
\usepackage{orcidlink}

\newtcblisting{codebox}{
  colback=blue!5!white,
  colframe=blue!75!black,
  boxrule=0.8pt,
  arc=4pt,
  listing only,
  listing options={
    basicstyle=\ttfamily\small,
    breaklines=true,
    postbreak=\mbox{\textcolor{red}{$\hookrightarrow$}\space},
  },
  enhanced,
  breakable,
  left=1mm, right=1mm,
  top=1mm, bottom=1mm,
  boxsep=1mm
}

\newtcblisting{terminalbox}{
  colback=yellow!6!orange!5!white,
  colframe=yellow!40!orange!50!black,
  boxrule=0.8pt,
  arc=4pt,
  listing only,
  listing options={
    basicstyle=\ttfamily\small,
    breaklines=true,
    columns=fullflexible,
    keepspaces=true,
  },
  enhanced,
  breakable,
  left=1mm, right=1mm,
  top=1mm, bottom=1mm,
  boxsep=1mm
}

\begin{document}

\title{\bf{\textsc{RooAgent}: An LLM Agent for \Root-Based High Energy Physics Analysis}}

\author{Aman Desai \orcidlink{0000-0003-2631-9696}\\[4pt]
\small Department of Physics, Adelaide University, Adelaide, SA 5005, Australia\\
\small\texttt{aman.desai@adelaide.edu.au}}

\date{\today}

\maketitle

\begin{abstract}
We present \rooagent as a natural-language interface for \Root-based high energy physics data analysis. The package provides physics analysis functions as tools that an LLM agent invokes in response to plain-language prompts. Two operating modes are supported: a \langraph-based agent compatible with \OpenAI's \GPTfourone via GitHub Copilot and with \DeepSeekVthree via \Ollama, and a Model Context Protocol server for use with the Anthropic \Claude CLI (\ClaudeSonnet). In both modes the analysis logic is implemented in \pyroot and the LLM selects tools and supplies the required arguments. The package supports histogram inspection, event selection, visualisation of kinematic distributions, fitting, and significance estimation, among other tasks. We illustrate \rooagent with tests based on Monte Carlo simulations of $pp\to ZH$ ($Z\to\ell^+\ell^-$, $H\to b\bar{b}$), a multi-task signal-background workflow, a toy statistical analysis, and an application to ATLAS open data for $H\to ZZ^*\to 4\ell$. The package is available on \pypi and the source code is hosted at \url{https://github.com/amanmdesai/RooAgent}.
\end{abstract}

\section{Introduction}\label{sec:intro}

High Energy Physics (HEP) data analysis relies on the \Root framework~\cite{Brun:1997pa} for tasks such as event selection, histogram preparation, and statistical inference in collider physics studies. \Root and its Python interface \pyroot~\cite{Galli:2020boj} provide a flexible toolkit, but working with them requires familiarity with their API conventions\footnote{API: Application Programming Interface} and the structure of each input sample. A natural-language interface that maps plain-language goals to \Root function calls can benefit new users and streamline routine tasks.

Large language models (LLMs)~\cite{zhao2026surveylargelanguagemodels} can invoke external functions through the so-called `tool calls'~\cite{schick2023toolformer,parisi2022talm}, enabling multi-step HEP workflows from natural-language prompts. Machine learning has been widely used in HEP data analysis~\cite{Radovic:2018dip,Guest:2018yhq,Albertsson:2018maf,Feickert:2021ajf,Plehn:2022ftl}, and LLM-based agents for tasks such as event selection, histogram production, statistical inference, and result interpretation~\cite{Jiao:2026dsu,Zhang:2024kws,Badea:2026klb,Aarrestad:2026xrs,Gendreau-Distler:2025fsj,Diefenbacher:2025zzn,Menzo:2025cim,Hill:2026naa,Esmail:2026jpb,Lo:2026use,Saito:2026tfq,Birk:2026zpd,Moreno:2026mqk,He:2026drsai}, and automated event generation~\cite{Plehn:2026gxv}, among other aspects have been actively studied.

\rooagent~\cite{aman_desai_2026_20249499} provides a library of analysis tools invoked by an LLM agent. The package wraps \pyroot~\cite{Galli:2020boj} functions as tools. The LLM interprets the user's input, selects the appropriate tools, and supplies the required arguments, while \Root performs the computation. The package supports two operating modes: a \langraph~\cite{langgraph2024} based agent for use with the \OpenAI~\cite{openai2023gpt4,githubcopilot2025} or \Ollama~\cite{ollama2025} APIs and a Model Context Protocol (MCP)~\cite{AnthropicMCP} server for use with the \Claude CLI~\cite{anthropic2024claude}. In both modes the tool set aims to cover common tasks in \Root-based analyses: file and tree inspection, histogram filling and visualisation, event counting and cutflow generation, significance calculation, parametric fitting, and data export. 

This paper is organised as follows. Section~\ref{sec:software} describes the installation, operating modes, and tool set of the \rooagent package. Section~\ref{sec:results} presents a series of benchmark tests performed on Monte Carlo simulated $ZH$ events, covering file inspection, histogram plotting, event counting, variable definition, fitting, cut optimisation, and a multi-task signal-background analysis. Section~\ref{sec:stats} applies \rooagent to a toy statistical analysis illustrating signal discovery and exclusion workflows. Section~\ref{sec:opendata} shows \rooagent applied to ATLAS open data for the $H\to ZZ^*\to 4\ell$ channel. Section~\ref{sec:conclusions} summarises the work.

\section{\rooagent Software}\label{sec:software}

\rooagent is written in Python and supports two operating modes: a \langraph-based agent mode for use with \OpenAI or \Ollama models and a Model Context Protocol (MCP) server mode for use with \Claude CLI. Both modes expose the same underlying \Root analysis tools, implemented in \pyroot~\cite{Galli:2020boj}. In both cases, \Root~\cite{Brun:1997pa} must be installed separately.

\subsection{\langraph Agent Mode}\label{sec:software:langgraph}

In this mode, the LLM reasons over user prompts, selects tools, constructs arguments, and calls \pyroot functions iteratively until the goal is met. Installation is via \pypi\footnote{\pypi version of the software supports access to LLMs via GitHub models only.}:

\begin{terminalbox}
pip install rooagent
\end{terminalbox}

or from source:

\begin{terminalbox}
git clone https://github.com/amanmdesai/RooAgent.git
cd RooAgent
pip install .
\end{terminalbox}

The additional required Python packages are:
\begin{itemize}
  \item \texttt{langchain}~\cite{langchain2022}, \texttt{langchain-core}, \texttt{langgraph}~\cite{langgraph2024}
  \item either \texttt{langchain-openai} or \texttt{langchain-ollama}~\cite{chatollama2025}
  \item \texttt{pandas}~\cite{McKinney:2010pandas}, \texttt{numpy}~\cite{Harris:2020numpy}, \texttt{scipy}~\cite{Virtanen:2020scipy}, \texttt{matplotlib}~\cite{Hunter:2007matplotlib}
\end{itemize}

By default, the package uses \OpenAI's \GPTfourone model~\cite{openai2023gpt4} accessed via GitHub Copilot's API~\cite{githubcopilot2025}. Runtime behaviour is controlled through the following environment variables:
\begin{itemize}
  \item \texttt{GITHUB\_TOKEN}: authentication token for model access.
  \item \texttt{MODEL} (optional): model identifier. Default is \texttt{openai/gpt-4.1}.
  \item \texttt{ROOAGENT\_SEED} (optional): random seed for reproducibility. Default is \texttt{7}.
\end{itemize}

To use locally hosted models, switch to the \Ollama branch:

\begin{terminalbox}
git clone https://github.com/amanmdesai/RooAgent.git
cd RooAgent
git checkout ollama_models
pip install .
\end{terminalbox}

This configures \langraph~\cite{langgraph2024} to use the \DeepSeekVthree model~\cite{DeepSeek2024v3} via \Ollama~\cite{ollama2025}. Any other locally installed \Ollama model can also be used by configuring \texttt{MODEL} parameter in the terminal.

The tool set is identical across both backends. The system's prompt is stored in \texttt{agent.py}.

\subsection{Model Context Protocol Mode}\label{sec:software:mcp}

\rooagent can also run as an MCP server integrated with the \Claude CLI~\cite{AnthropicMCP}. Here, \Claude acts as both the LLM and orchestration layer, with no \langchain or \langraph installation required. The MCP variant is available on the \texttt{claude\_models} branch and can be installed using the following commands:

\begin{terminalbox}
git clone https://github.com/amanmdesai/RooAgent.git
cd RooAgent
git checkout claude_models
pip install .
\end{terminalbox}

Additional required packages for this mode:
\begin{itemize}
  \item \texttt{fastmcp}~\cite{fastmcp2024}: the \texttt{FastMCP} library used to implement the \textsc{MCP} server
  \item \texttt{pandas}~\cite{McKinney:2010pandas}, \texttt{numpy}~\cite{Harris:2020numpy}, \texttt{scipy}~\cite{Virtanen:2020scipy}, \texttt{matplotlib}~\cite{Hunter:2007matplotlib}
\end{itemize}

After installation, register the server once with the \Claude CLI:

\begin{terminalbox}
claude mcp add rooagent --rooagent-mcp
\end{terminalbox}

To run an analysis, navigate to the directory containing the \Root files and launch \Claude:

\begin{terminalbox}
cd /path/to/your/root/files
claude
\end{terminalbox}

The repository's \texttt{CLAUDE.md} is read automatically at the start of \Claude session and it carries the system's prompt — including available tools, recommended workflows, default parameters, and analysis rules — so no additional configuration is needed. The model chosen as default is the one configured in the user's \Claude settings. For tests here, we used \ClaudeSonnet~\cite{anthropic2024claude}.

\subsection{Tool Set and Internal Architecture}\label{sec:software:tools}

\begin{table}[h]
\centering
\caption{Complete tool set exposed by \rooagent in both operating modes. All tools operate on \Root files via \pyroot and return structured text output; plotting tools additionally produce PDF figures.}
\footnotesize
\begin{tabular}{lll}
\toprule
\textbf{Category} & \textbf{Tool} & \textbf{Description} \\
\midrule
\multirow{1}{*}{Inspection}
  & \texttt{inspect\_root\_data}      & List file contents, TTree structure, and branch names/types \\
\midrule
\multirow{4}{*}{Counting}
  & \texttt{apply\_cut\_and\_count}   & Count (un)weighted events passing a boolean selection \\
  & \texttt{generate\_cutflow}        & Sequential cutflow with weighted yields for multiple files \\
  & \texttt{compute\_significance}    & $S/\sqrt{S+B}$ for arbitrary selection cuts \\
  & \texttt{compute\_efficiency}      & Selection efficiency relative to a denominator cut \\
\midrule
\multirow{3}{*}{Statistics}
  & \texttt{histogram\_significance\_and\_cls} & $Z$ and CL$_s$ from stored TH1 histograms \\
  & \texttt{summarize\_parameter\_scan}        & Rank a scan of parameter points by significance \\
  & \texttt{plot\_significance\_and\_cls}      & Plot $Z$ or CL$_s$ vs.\ a scan parameter \\
\midrule
\multirow{3}{*}{Histograms}
  & \texttt{histogram\_integral}      & Integral of a stored TH1 in a specified $x$-range \\
  & \texttt{get\_histogram\_stats}    & Mean, RMS, and entries of a stored TH1 \\
  & \texttt{root\_tree\_to\_histogram}& Fill and save a TH1 from TTree branches with optional cuts \\
\midrule
\multirow{2}{*}{Plotting}
  & \texttt{plot}                     & Single or overlay/stacked histogram from TH1s or TTree branches \\
  & \texttt{plot\_2d}                 & Two-dimensional histogram from two TTree branches \\
\midrule
\multirow{1}{*}{Fitting}
  & \texttt{fit\_distribution}        & Gaussian, exponential, or polynomial fit to a TH1 or TTree variable \\
\midrule
\multirow{3}{*}{Variables}
  & \texttt{define\_variable}         & Compute a derived branch and write to a new ROOT file \\
  & \texttt{define\_variable\_and\_plot} & Define branch, apply cuts, and plot in one step \\
  & \texttt{find\_optimal\_cut}       & Scan a threshold and return the cut maximising $S/\sqrt{S+B}$ \\
\midrule
\multirow{1}{*}{Export}
  & \texttt{root\_tree\_to\_csv}      & Export selected TTree branches to a CSV file \\
\bottomrule
\end{tabular}
\label{tab:tools}
\end{table}

\begin{figure*}[p]
    \centering
    \includegraphics[angle=90,origin=c,width=0.95\textheight,height=0.95\textwidth,keepaspectratio]{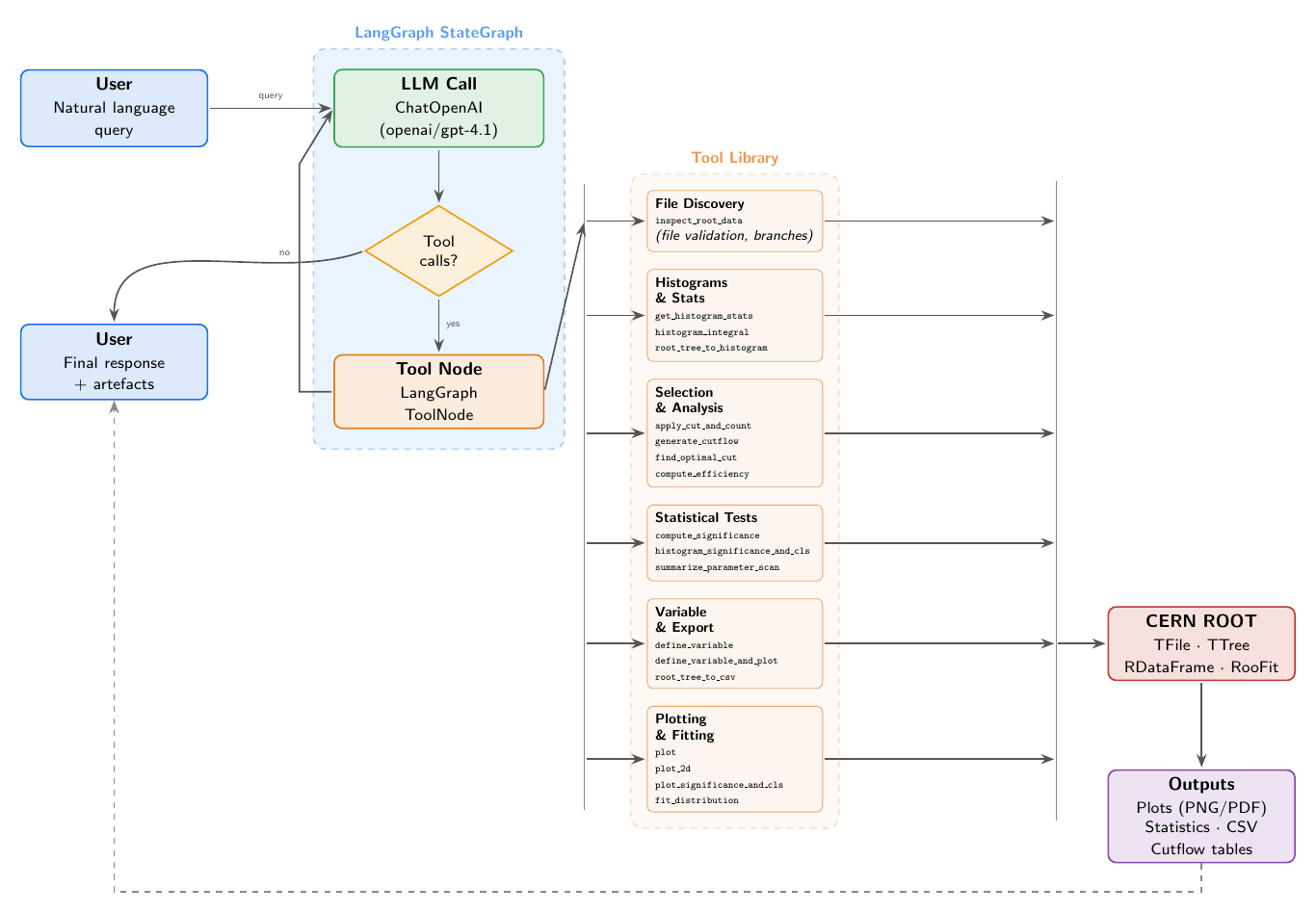}
    \caption{Architecture of \rooagent in the \langraph-based operating mode. A user issues natural-language queries to an LLM-based agent, which selects and invokes the appropriate \Root analysis tools and returns results as plots, statistics, tables, or summaries. The system supports iterative reasoning: tools may be called multiple times.}
    \label{fig:workflow}
\end{figure*}

The \langraph-based architecture of \rooagent is illustrated in \autoref{fig:workflow}. In the \langraph mode, the tools are registered with the \texttt{langchain\_core} \texttt{@tool} decorator and called by the agent. In the MCP mode, the same functions are registered directly with the \texttt{FastMCP}~\cite{fastmcp2024} server. \pyroot handles the computation in both cases, with \texttt{RDataFrame}~\cite{Piparo:2019bnl} for variable definition and histogram construction and \textsc{RooFit}~\cite{Verkerke:2003ir} for likelihood-based fitting. Statistical inference uses \texttt{scipy}~\cite{Virtanen:2020scipy}; visualisation uses \Root's canvas routines or \texttt{matplotlib}~\cite{Hunter:2007matplotlib}.

Both modes include a system's prompt with the complete tool registry, recommended workflows, and default parameters. In \langraph mode it is embedded in \texttt{agent.py} whereas in MCP mode it is stored in \texttt{CLAUDE.md} and is read automatically by the \Claude CLI at session start.  \autoref{tab:tools} lists each tool with its category and its utility.

\section{Results with \rooagent}\label{sec:results}

We evaluate \rooagent on Monte Carlo simulated samples. In each benchmark test, the user provides a natural-language prompt. The LLM selects the appropriate tools, infers the required arguments, and produces the final output without human intervention.

\subsection{Monte Carlo Sample}\label{sec:mc}

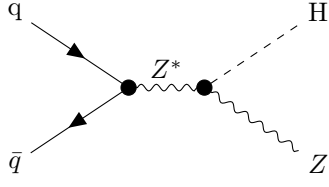
\begin{figure}[h!]
    \centering
    \begin{tikzpicture}
\begin{feynman}

    \vertex (q1) at (0.5,1) {q};
    \vertex (q2) at (0.5,-1) {\(\bar{q}\)};

    \vertex [dot, minimum size=2mm] (v1) at (2,0) {};

    \vertex [dot, minimum size=2mm] (v2) at (3,0) {};

    \vertex (h) at (4.5,1) {H};
    \vertex (z) at (4.5,-1) {$Z$};

    \diagram* {

        (q1) -- [fermion] (v1) -- [fermion] (q2),

        (v1) -- [boson, edge label=$Z^*$] (v2),

        (v2) -- [boson] (z),
        (v2) -- [scalar] (h),
    };
\end{feynman}
\end{tikzpicture}
    \caption{Representative leading-order Feynman diagram for $q\bar{q}\to ZH$.}
    \label{fig:ZH_diagram}
\end{figure}

Development and testing used two Monte Carlo samples. The signal sample corresponds to $pp \to Z(\to \ell^+\ell^-)H(\to b\bar{b})$; a representative leading-order diagram is shown in \autoref{fig:ZH_diagram}. The background sample is the continuum process $pp \to \ell^+\ell^- b\bar{b}$, with the same visible final state but no intermediate Higgs boson. We simulated $500\,000$ events for each process using \MG~\cite{Alwall:2014hca} at the LHC Run~3 centre-of-mass energy of $\sqrt{s} = 13.6$\,\TeV. The PDF4LHC21 PDF set~\cite{PDF4LHCWorkingGroup:2022cjn} was used, accessed via \textsc{Lhapdf6}~\cite{Buckley:2014ana}. The hard-scatter events then passed via \Pythia~\cite{Bierlich:2022pfr} to incorporate effects of parton shower and hadronisation. The resulting hadrons were clustered into jets with \FastJet~\cite{Cacciari:2011ma} via the \MadAnalysis~\cite{Conte:2018vmg} interface, using the anti-$k_T$ algorithm~\cite{Cacciari:2008gp} with $R = 0.4$. The \MadAnalysis output in the \lhe format~\cite{Alwall:2006yp} was processed with the \lhereader~\cite{Desai:2026lst} and an \event function~\cite{Desai:2026lst} to produce a \Root file with a \texttt{TTree} of kinematic branches and pre-filled \texttt{TH1D} histograms.

The \event function applies object selection and reconstructs composite final states from four-momenta. Electrons, muons, and jets are required to satisfy $p_T > 25$\,\gev and $|\eta| < 2.5$. A lepton--jet isolation requirement $\Delta R(\ell, j) > 0.4$ is imposed. The dilepton invariant mass is required to satisfy $m_{\ell\ell}\in(76, 106)$\,\gev, consistent with an on-shell $Z$ boson. $b$-jets are identified using generator-level particle flavour information. Each event carries a weight combining the process cross-section, generator weight, and a fixed normalisation factor, used throughout to produce yields at a defined integrated luminosity.

The tests below use these two \Root files. The signal $m_{bb}$ distribution peaks broadly near $108$--$110$\,\gev; the background spectrum is approximately exponential. File contents are summarised in \autoref{tab:struct1} and \autoref{tab:struct2}.

\begin{table}[h]
\centering
\footnotesize
\begin{minipage}[t]{0.4\textwidth}
\centering
\caption{Summary of one-dimensional histograms stored in the \Root file.}
\label{tab:struct1}
\begin{tabular}{ll}
\toprule
Histogram Name & Description \\
\midrule
MET           & Missing transverse energy \\
bbEta         & Selected bb-system pseudorapidity \\
bbPt          & Selected bb-system transverse momentum \\
bjet1Pt       & Leading b-jet transverse momentum \\
bjet2Pt       & Subleading b-jet transverse momentum \\
electronIso   & Electron isolation \\
electronPt    & Electron transverse momentum \\
jetPt         & b-jet transverse momentum \\
m\_bb         & Selected b-jet pair mass \\
m\_ee         & Dielectron invariant mass \\
m\_jj         & Dijet invariant mass \\
m\_mumu       & Dimuon invariant mass \\
muonIso       & Muon isolation \\
muonPt        & Muon transverse momentum \\
n\_bjets      & b-jet multiplicity \\
n\_cleanJets  & Number of isolated jets \\
n\_gluonJets  & Number of gluon-initiated jets \\
\bottomrule
\end{tabular}
\end{minipage}
\hfill
\begin{minipage}[t]{0.4\textwidth}
\centering
\caption{Branches stored in the \texttt{events} TTree.}
\label{tab:struct2}
\begin{tabular}{lll}
\toprule
Branch Name & Type & Description \\
\midrule
electron\_pt     & vector<float> & $p_T^{e}$ \\
electron\_eta    & vector<float> & $\eta^{e}$ \\
electron\_phi    & vector<float> & $\phi^{e}$ \\
electron\_charge & vector<int>   & $q^{e}$ \\
muon\_pt         & vector<float> & $p_T^{\mu}$ \\
muon\_eta        & vector<float> & $\eta^{\mu}$ \\
muon\_phi        & vector<float> & $\phi^{\mu}$ \\
muon\_charge     & vector<int>   & $q^{\mu}$ \\
jet\_pt          & vector<float> & $p_T^{\mathrm{jet}}$ \\
jet\_eta         & vector<float> & $\eta^{\mathrm{jet}}$ \\
jet\_phi         & vector<float> & $\phi^{\mathrm{jet}}$ \\
jet\_flavour     & vector<int>   & $f_{\mathrm{jet}}$ \\
met\_pt          & float         & $E_T^{\mathrm{miss}}$ \\
met\_phi         & float         & $\phi(E_T^{\mathrm{miss}})$ \\
m\_bb            & float         & $m_{bb}$ \\
bbPt             & float         & $p_T^{bb}$ \\
bbEta            & float         & $\eta^{bb}$ \\
dilep\_pt        & float         & $p_T^{\ell\ell}$ \\
mll              & float         & $m_{\ell\ell}$ \\
weight           & float         & Per-event weight \\
\bottomrule
\end{tabular}
\end{minipage}
\end{table}

\clearpage
\subsection{Performance of \rooagent}\label{sec:performance}

In the following, all output boxes (``Final Answer'') were generated entirely by LLMs. Human involvement was limited to writing the input prompt. Final results were post-processed by \Claude (\ClaudeSonnet)\footnote{irrespective of the initial LLM model used for testing}, e.g.\ fixing latex issues or merging multiple samples' cutflows. No further review was applied beyond the tools' internal checks. All tests used \GPTfourone~\cite{openai2023gpt4} via GitHub Copilot~\cite{githubcopilot2025}, unless stated otherwise. Qualitatively, similar results were obtained with \DeepSeekVthree~\cite{DeepSeek2024v3} via \Ollama~\cite{ollama2025}.

\subsection{Tests with Single Task Prompts}

In \autoref{tab:benchmark} we present a list of test that were carried out using a single task prompt. A check mark indicates the \rooagent produced a result, not necessarily a correct one. Subsections below give the prompt and response for each test.

\begin{table}[h]
\centering
\caption{Summary of individual benchmark tests. Elapsed times include all LLM reasoning and tool-call round trips. Outcome: \checkmark~= agent-reported answer, $\times$~= agent-reported error. The multi-task test is listed separately.}
\small
\begin{tabular}{clllcc}
\toprule
\textbf{Test} & \textbf{Task} & \textbf{Category} & \textbf{Primary tools} & \textbf{Time (s)} & \textbf{Outcome} \\
\midrule
1 & File inspection              & Inspection  & \texttt{inspect\_root\_data}              &   7.46 & \checkmark \\
2 & Branch listing               & Inspection  & \texttt{inspect\_root\_data}              &   9.73 & \checkmark \\
3 & Branch data types            & Inspection  & \texttt{inspect\_root\_data}              &   9.82 & \checkmark \\
4 & Histogram statistics         & Histograms  & \texttt{get\_histogram\_stats}            &   8.37 & \checkmark \\
5 & Histogram (log scale)        & Plotting    & \texttt{plot}                             &  20.66 & \checkmark \\
6 & Normalised histogram         & Plotting    & \texttt{plot}                             &  18.27 & \checkmark \\
7 & Tree variable plot           & Plotting    & \texttt{root\_tree\_to\_histogram, plot}  &  12.69 & \checkmark \\
9 & 2D histogram                 & Plotting    & \texttt{plot\_2d}                         &  20.50 & \checkmark \\
10 & 2D correlation               & Plotting    & \texttt{plot\_2d}                         &  16.89 & \checkmark \\
11 & Event count (scalar)         & Counting    & \texttt{apply\_cut\_and\_count}           &   8.19 & $\times$   \\
12 & Event count (vector, all)    & Counting    & \texttt{apply\_cut\_and\_count}           &  20.07 & \checkmark \\
13 & Event count (vector, any)    & Counting    & \texttt{apply\_cut\_and\_count}           &  80.28 & \checkmark \\
14 & Significance with cut        & Counting    & \texttt{compute\_significance}            &  20.48 & \checkmark \\
15 & Cut scan (lower bound)       & Variables   & \texttt{find\_optimal\_cut}               & 134.28 & \checkmark \\
16 & Variable definition          & Variables   & \texttt{define\_variable}                 &  17.27 & \checkmark \\
17 & Gaussian fit                 & Fitting     & \texttt{fit\_distribution}                &   8.19 & \checkmark \\
19 & CSV export                   & Export      & \texttt{root\_tree\_to\_csv}              &  37.63 & \checkmark \\
20 & Overlay plot ($m_{bb}$)      & Plotting    & \texttt{plot}                             &  20.58 & \checkmark \\
21 & Overlay plot ($\mu$ $p_T$)   & Plotting    & \texttt{plot}                             &  23.65 & \checkmark \\
22  & Cut scan (bounded range)     & Variables   & \texttt{find\_optimal\_cut}               & 178.84 & \checkmark \\
\bottomrule
\end{tabular}
\label{tab:benchmark}
\end{table}

\subsubsection*{Test 1: Signal File Inspection}

\testprompt{Tell me what is stored inside \texttt{signal.root}.

\textit{Output requirements:} Save all generated PDF files under \texttt{output}; every PDF filename must end with \texttt{\_test01.pdf}.}

\testelapsed{7.46\,s}

\testanswer{The file \texttt{signal.root} contains the following objects:
\begin{itemize}[nosep]
  \item \textbf{17 histograms (TH1F):} \texttt{MET}, \texttt{bbEta}, \texttt{bbPt}, \texttt{bjet1Pt}, \texttt{bjet2Pt}, \texttt{electronIso}, \texttt{electronPt}, \texttt{jetPt}, \texttt{m\_bb}, \texttt{m\_ee}, \texttt{m\_jj}, \texttt{m\_mumu}, \texttt{muonIso}, \texttt{muonPt}, \texttt{n\_bjets}, \texttt{n\_cleanJets}, \texttt{n\_gluonJets}
  \item \textbf{1 TTree:} \texttt{events}
\end{itemize}
No PDF files were generated for this test, as only a content listing was requested.}

\subsubsection*{Test 2: Signal Tree Branch Listing}

\testprompt{List the branches in the \texttt{events} tree from \texttt{signal.root}.

\textit{Output requirements:} Save all generated PDF files under \texttt{output}; every PDF filename must end with \texttt{\_test02.pdf}.}

\testelapsed{9.73\,s}

\testanswer{The branches in the \texttt{events} tree from \texttt{signal.root} are:
\begin{itemize}[nosep,label=\textbullet]
  \item \texttt{electron\_pt}, \texttt{electron\_eta}, \texttt{electron\_phi}, \texttt{electron\_charge} : \texttt{vector<float>} / \texttt{vector<int>}
  \item \texttt{muon\_pt}, \texttt{muon\_eta}, \texttt{muon\_phi}, \texttt{muon\_charge} : \texttt{vector<float>} / \texttt{vector<int>}
  \item \texttt{dilep\_pt}, \texttt{mll}, \texttt{met\_pt}, \texttt{met\_phi}, \texttt{m\_bb}, \texttt{bbPt}, \texttt{bbEta}, \texttt{weight} : \texttt{Float\_t}
  \item \texttt{jet\_pt}, \texttt{jet\_eta}, \texttt{jet\_phi} : \texttt{vector<float>}
  \item \texttt{jet\_flavour} : \texttt{vector<int>}
\end{itemize}
No PDF files were generated.}

\subsubsection*{Test 3: Background Tree Branch Data Types}

\testprompt{Show the branch data types for the \texttt{events} tree in \texttt{background.root}.

\textit{Output requirements:} Save all generated PDF files under \texttt{output}; every PDF filename must end with \texttt{\_test03.pdf}.}

\testelapsed{9.82\,s}

\testanswer{The branch data types for \texttt{background.root} are identical to those in \texttt{signal.root}:
\begin{itemize}[nosep,label=\textbullet]
  \item \texttt{electron\_pt}, \texttt{electron\_eta}, \texttt{electron\_phi} : \texttt{vector<float>}
  \item \texttt{electron\_charge}, \texttt{muon\_charge}, \texttt{jet\_flavour} : \texttt{vector<int>}
  \item \texttt{muon\_pt}, \texttt{muon\_eta}, \texttt{muon\_phi}, \texttt{jet\_pt}, \texttt{jet\_eta}, \texttt{jet\_phi} : \texttt{vector<float>}
  \item \texttt{dilep\_pt}, \texttt{mll}, \texttt{met\_pt}, \texttt{met\_phi}, \texttt{m\_bb}, \texttt{bbPt}, \texttt{bbEta}, \texttt{weight} : \texttt{Float\_t}
\end{itemize}
No PDF files were generated.}

\subsubsection*{Test 4: m\_bb Histogram Statistics}

\testprompt{Give statistics for the \texttt{m\_bb} histogram in \texttt{signal.root}.

\textit{Output requirements:} Save all generated PDF files under \texttt{output}; every PDF filename must end with \texttt{\_test04.pdf}.}

\testelapsed{8.37\,s}

\testanswer{Statistics for the \texttt{m\_bb} histogram in \texttt{signal.root}:
\begin{center}
\begin{tabular}{ll}
\toprule
\textbf{Statistic} & \textbf{Value} \\
\midrule
Mean    & 107.27 \\
RMS     & 18.03  \\
Entries & 82\,645 \\
\bottomrule
\end{tabular}
\end{center}
No PDF was generated.}

\subsubsection*{Test 5: m\_bb Histogram with Log Scale}

\testprompt{Draw the \texttt{m\_bb} histogram from \texttt{signal.root} with a logarithmic y-axis and save it.

\textit{Output requirements:} Save all generated PDF files under \texttt{output}; every PDF filename must end with \texttt{\_test05.pdf}.}

\testelapsed{20.66\,s}

\testanswer{The \texttt{m\_bb} histogram from \texttt{signal.root} has been drawn with a logarithmic y-axis and saved as \texttt{output/m\_bb\_test05.pdf}.

\medskip\centering
\includegraphics[width=0.40\textwidth]{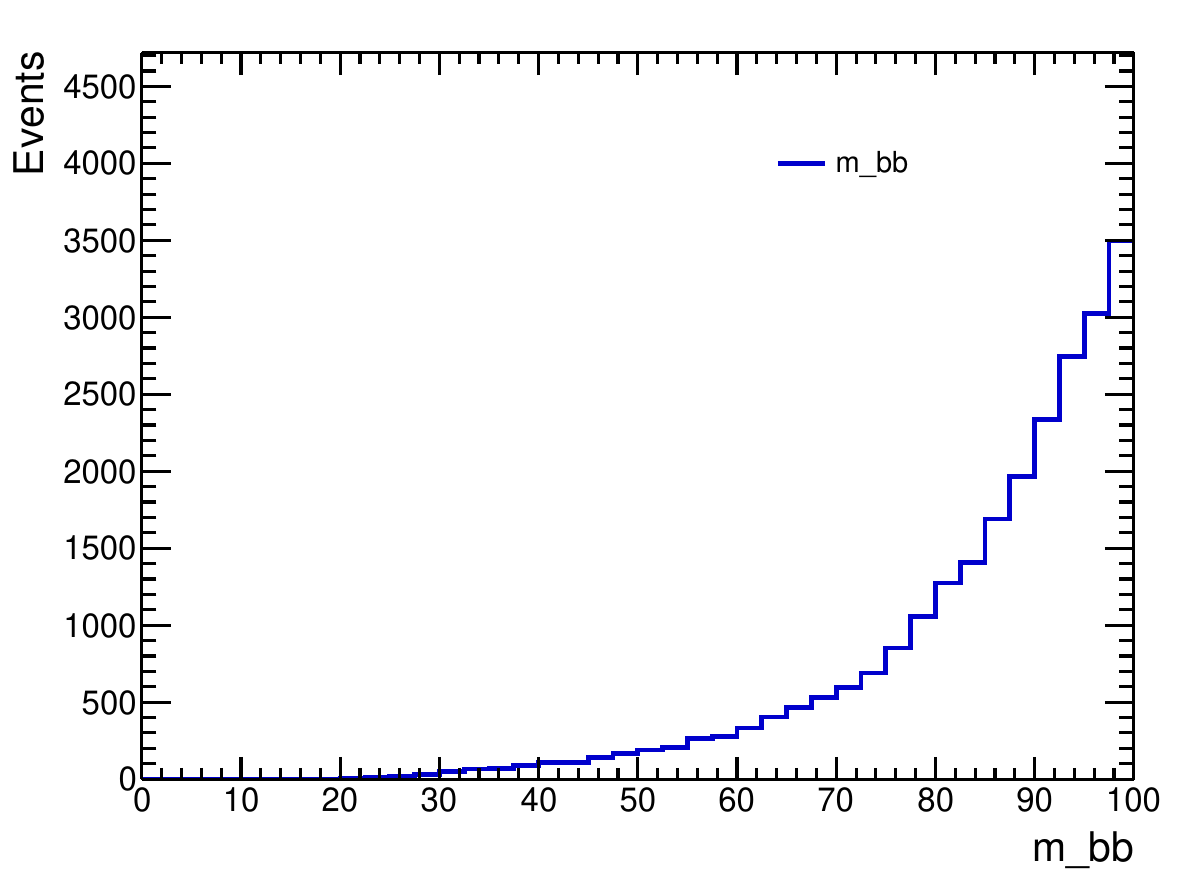}
\captionof{figure}{$m_{bb}$ histogram from \texttt{signal.root} with a logarithmic $y$-axis. Plots generated by \rooagent.}}

\subsubsection*{Test 6: Normalised mll Histogram}

\testprompt{Plot the \texttt{mll} histogram from \texttt{background.root} with normalised area and save it.

\textit{Output requirements:} Save all generated PDF files under \texttt{output}; every PDF filename must end with \texttt{\_test06.pdf}.}

\testelapsed{18.27\,s}

\testanswer{The normalised \texttt{mll} histogram from \texttt{background.root} has been plotted (area normalised to 1) and saved as \texttt{output/mll\_background\_test06.pdf}.

\medskip\centering
\includegraphics[width=0.40\textwidth]{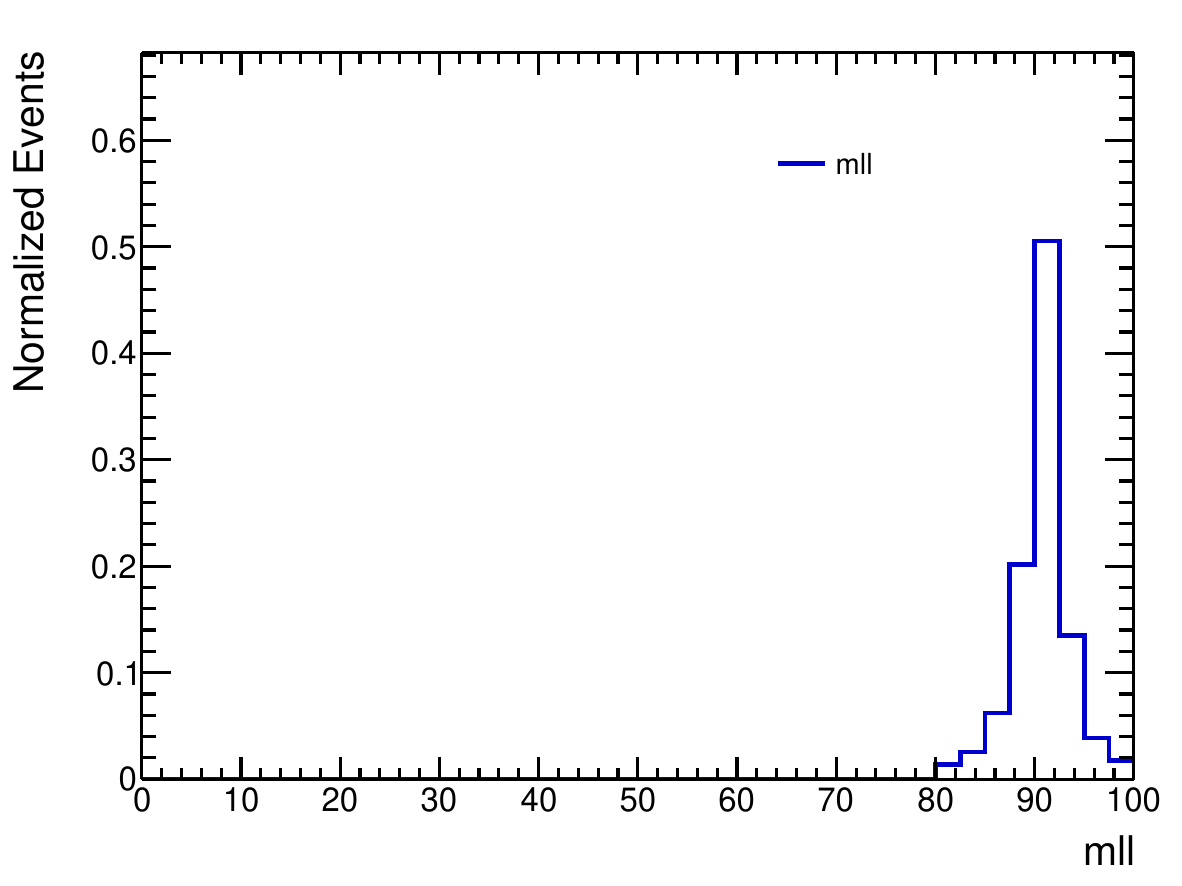}
\captionof{figure}{Area-normalised $m_{\ell\ell}$ distribution from \texttt{background.root}. Plots generated by \rooagent.}}

\subsubsection*{Test 7: muon\_pt Distribution}

\testprompt{Show a plot of \texttt{muon\_pt} from the \texttt{events} tree in \texttt{background.root}.

\textit{Output requirements:} Save all generated PDF files under \texttt{output}; every PDF filename must end with \texttt{\_test07.pdf}.}

\testelapsed{12.69\,s}

\testanswer{The plot of \texttt{muon\_pt} from the \texttt{events} tree in \texttt{background.root} has been generated and saved as \texttt{output/muon\_pt\_test07.pdf}.

\medskip\centering
\includegraphics[width=0.40\textwidth]{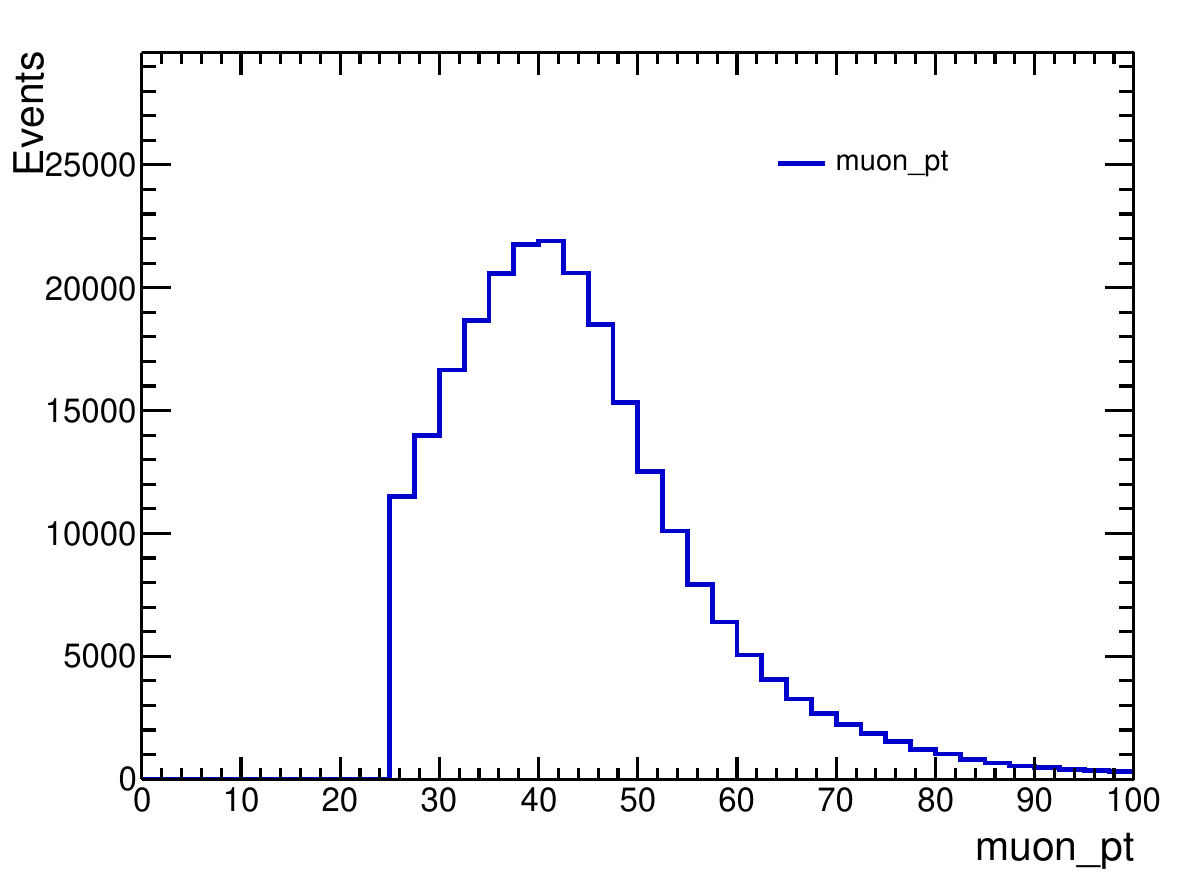}
\captionof{figure}{Muon $p_T$ distribution from the \texttt{events} tree in \texttt{background.root}. Plots generated by \rooagent.}}

\subsubsection*{Test 9: 2D Plot: jet\_pt vs.\ jet\_eta}

\testprompt{Draw a 2D plot of \texttt{jet\_pt} versus \texttt{jet\_eta} from \texttt{signal.root} using \texttt{jet\_eta} $\in [-4,4]$ and \texttt{jet\_pt} $\in [0,100]$.

\textit{Output requirements:} Save all generated PDF files under \texttt{output}; every PDF filename must end with \texttt{\_test09.pdf}.}

\testelapsed{20.50\,s}

\testanswer{The 2D plot of \texttt{jet\_pt} vs.\ \texttt{jet\_eta} from \texttt{signal.root} (\texttt{events} tree) has been generated with 40 bins in each direction and saved as \texttt{output/jet\_eta\_vs\_jet\_pt\_test09.pdf}.

\medskip\centering
\includegraphics[width=0.40\textwidth]{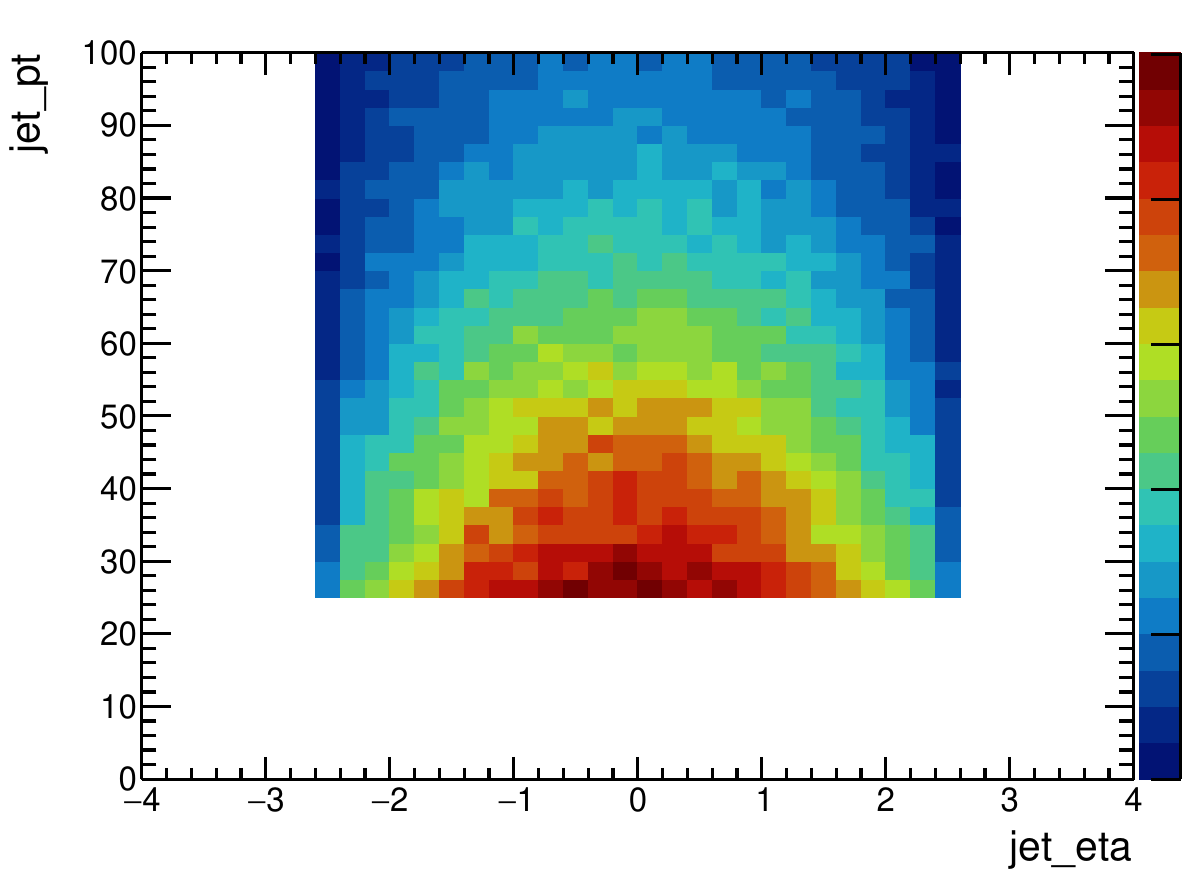}
\captionof{figure}{2D histogram of jet $p_T$ vs.\ jet $\eta$ from the \texttt{events} tree in \texttt{signal.root}. The $\eta$ range is $[-4,4]$ and the $p_T$ range is $[0,100]$\,\gev, each with 40 bins. Plots generated by \rooagent.}}

\subsubsection*{Test 10: Correlation: jet\_pt vs.\ dilep\_pt}

\testprompt{Show the correlation between \texttt{jet\_pt} and \texttt{dilep\_pt} in the \texttt{events} tree.

\textit{Output requirements:} Save all generated PDF files under \texttt{output}; every PDF filename must end with \texttt{\_test10.pdf}.}

\testelapsed{16.89\,s}

\testanswer{The correlation between \texttt{jet\_pt} and \texttt{dilep\_pt} has been visualised as a 2D histogram covering $[0,100]$\,\gev in both axes. Output saved as \texttt{output/signal\_jetpt\_dileppt\_test10.pdf}.

\medskip\centering
\includegraphics[width=0.40\textwidth]{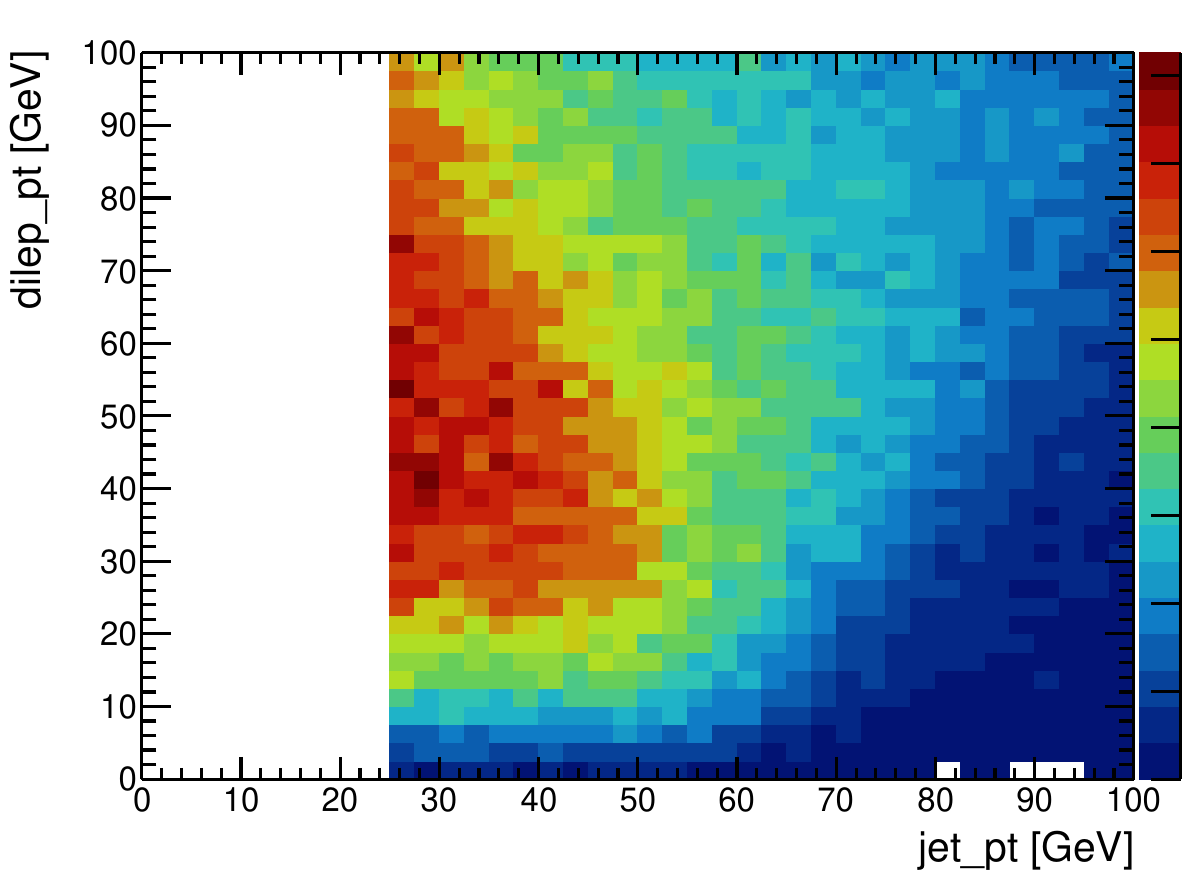}
\captionof{figure}{2D histogram of dilepton $p_T$ vs.\ jet $p_T$ for signal events, covering $[0,100]$\,\gev in each axis. Plots generated by \rooagent.}}

\subsubsection*{Test 11: Event Count: dilep\_pt $>$ 50}

\testprompt{Count events in \texttt{signal.root} where \texttt{dilep\_pt} $> 50$.

\textit{Output requirements:} Save all generated PDF files under \texttt{output}; every PDF filename must end with \texttt{\_test11.pdf}.}

\testelapsed{8.19\,s}

\testanswer{\textbf{Result:} The TTree named \texttt{Events} does not exist in \texttt{signal.root} (correct name is \texttt{events}). The agent requested clarification on the correct TTree name. No PDF was generated.

\textit{Note: this test revealed a case sensitivity issue in the tree name lookup.}}

\subsubsection*{Test 12: Event Count: All Jets with jet\_pt $>$ 20}

\testprompt{Count \texttt{signal.root} events where all jets have \texttt{jet\_pt} $> 20$.

\textit{Output requirements:} Save all generated PDF files under \texttt{output}; every PDF filename must end with \texttt{\_test12.pdf}.}

\testelapsed{20.07\,s}

\testanswer{\textbf{Result:} The number of events in \texttt{signal.root} where \emph{all} jets satisfy \texttt{jet\_pt} $> 20$ is \textbf{181\,932}. No PDF was generated.}

\subsubsection*{Test 13: Event Count: Any Jet with jet\_pt $>$ 30 and jet\_eta $<$ 2.5}

\testprompt{Count events in \texttt{signal.root} where any jet has \texttt{jet\_pt} $> 30$ and \texttt{jet\_eta} $< 2.5$.

\textit{Output requirements:} Save all generated PDF files under \texttt{output}; every PDF filename must end with \texttt{\_test13.pdf}.}

\testelapsed{80.28\,s}

\testanswer{\textbf{Result:} Events in \texttt{signal.root} where \emph{any} jet satisfies \texttt{jet\_pt} $> 30$ \emph{and} \texttt{jet\_eta} $< 2.5$: \textbf{171\,290}. No PDF was generated.}

\subsubsection*{Test 14: Significance S/$\sqrt{S+B}$ with m\_bb $>$ 120}

\testprompt{Compute $S/\sqrt{S+B}$ for signal and background with \texttt{m\_bb} $> 120$ using event weights.

\textit{Output requirements:} Save all generated PDF files under \texttt{output}; every PDF filename must end with \texttt{\_test14.pdf}.}

\testelapsed{20.48\,s}

\testanswer{Results for $S/\sqrt{S+B}$ with \texttt{m\_bb} $> 120$ and event weights:
\begin{center}
\begin{tabular}{ll}
\toprule
\textbf{Quantity}         & \textbf{Value} \\
\midrule
Signal yield ($S$)        & 796.54 \\
Background yield ($B$)    & 266\,959.81 \\
Significance $Z$          & 1.54 \\
\bottomrule
\end{tabular}
\end{center}
No PDF was generated.}

\subsubsection*{Test 15: m\_bb Cut Scan for Best Significance}

\testprompt{Scan the \texttt{m\_bb} cut to find the best significance and save the scan plot.

\textit{Output requirements:} Save all generated PDF files under \texttt{output}; every PDF filename must end with \texttt{\_test15.pdf}.}

\testelapsed{134.28\,s}

\testanswer{Scan summary:
\begin{center}
\begin{tabular}{ll}
\toprule
\textbf{Quantity}         & \textbf{Value} \\
\midrule
Optimal cut               & $m_{bb} > 75.0$\,\gev \\
Signal yield ($S$)        & 3\,244.7 \\
Background yield ($B$)    & 389\,298.2 \\
Best significance $Z$     & 5.18 \\
\bottomrule
\end{tabular}
\end{center}
Scan plot saved as \texttt{output/m\_bb\_scan\_test15.pdf}.

\medskip\centering
\includegraphics[width=0.40\textwidth]{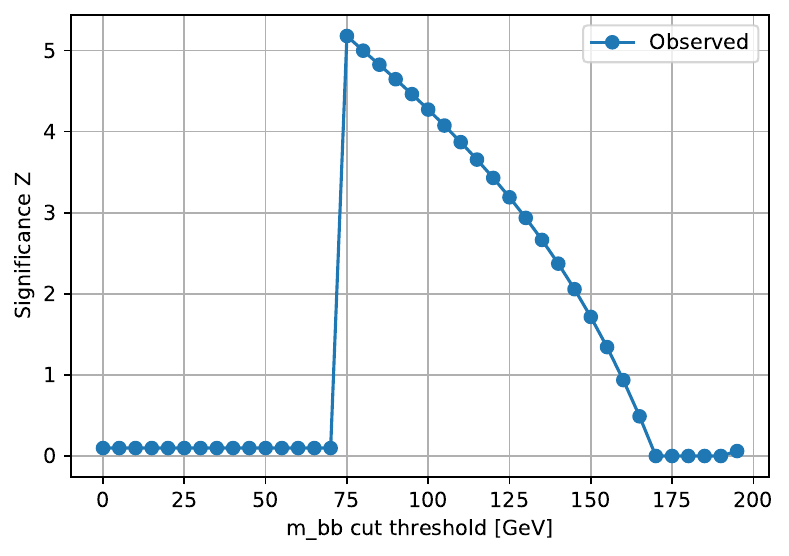}
\captionof{figure}{$S/\sqrt{S+B}$ significance as a function of the lower $m_{bb}$ threshold. The optimal cut is at $m_{bb}>75$\,\gev, giving $Z=5.18$. Plots generated by \rooagent.}}

\subsubsection*{Test 16: New Variable: leading\_jet\_pt}

\testprompt{Create \texttt{leading\_jet\_pt = jet\_pt[0]} in the \texttt{events} tree of \texttt{signal.root}.

\textit{Output requirements:} Save all generated PDF files under \texttt{output}; every PDF filename must end with \texttt{\_test16.pdf}.}

\testelapsed{17.27\,s}

\testanswer{The new variable \texttt{leading\_jet\_pt} (type \texttt{Float\_t}) has been successfully added to the \texttt{events} tree. The modified file was saved as \texttt{signal\_leadingjet\_test16.root}. All original branches remain unchanged. No PDF was generated.}

\subsubsection*{Test 17: Gaussian Fit to m\_bb}

\testprompt{Fit the \texttt{m\_bb} histogram from \texttt{signal.root} with a Gaussian.

\textit{Output requirements:} Save all generated PDF files under \texttt{output}; every PDF filename must end with \texttt{\_test17.pdf}.}

\testelapsed{8.19\,s}

\testanswer{Gaussian fit to the \texttt{m\_bb} histogram is complete:
\begin{center}
\begin{tabular}{ll}
\toprule
\textbf{Parameter} & \textbf{Value} \\
\midrule
Amplitude ($p_0$) & 404.89 \\
Mean ($\mu$)      & 108.32\,\gev \\
Sigma ($\sigma$)  & 13.23\,\gev \\
$\chi^2/\mathrm{ndf}$ & $12482.13 / 34$ \\
\bottomrule
\end{tabular}
\end{center}
Fit plot saved as \texttt{output/m\_bb\_fit\_test17.pdf}.

\medskip\centering
\includegraphics[width=0.40\textwidth]{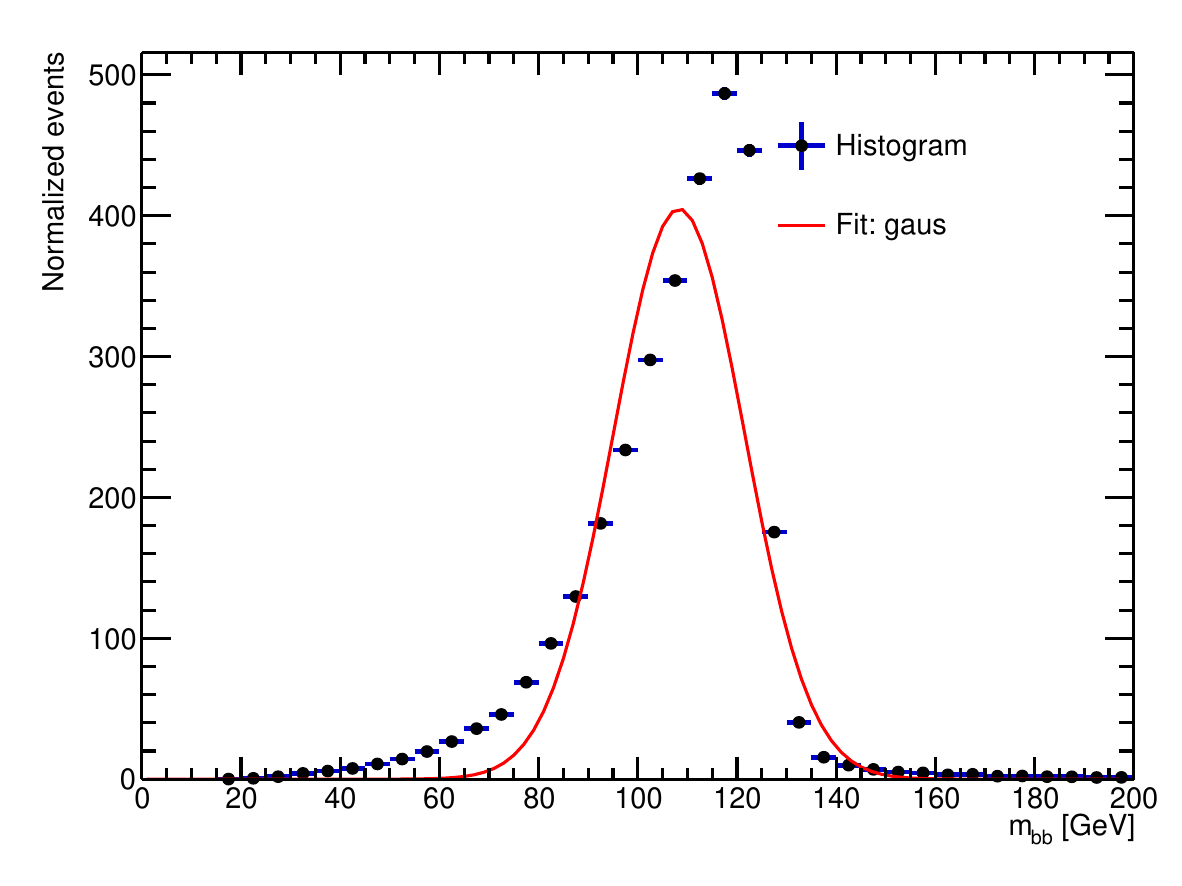}
\captionof{figure}{Gaussian fit to the $m_{bb}$ histogram from \texttt{signal.root}. The fitted parameters are $\mu = 108.32$\,\gev and $\sigma = 13.23$\,\gev. Plots generated by \rooagent.}}

\subsubsection*{Test 19: CSV Export of jet\_pt, jet\_eta, dilep\_pt}

\testprompt{Export \texttt{jet\_pt}, \texttt{jet\_eta}, and \texttt{dilep\_pt} from the \texttt{events} tree to CSV.

\textit{Output requirements:} Save all generated PDF files under \texttt{output}; every PDF filename must end with \texttt{\_test19.pdf}.}

\testelapsed{37.63\,s}

\testanswer{The branches \texttt{jet\_pt}, \texttt{jet\_eta}, and \texttt{dilep\_pt} from the \texttt{events} tree in \texttt{signal.root} have been exported to \texttt{output/signal\_events\_test19.csv}. No PDF was generated.}

\subsubsection*{Test 20: m\_bb Overlay: Signal vs.\ Background}

\testprompt{Overlay \texttt{m\_bb} from \texttt{signal.root} and \texttt{background.root} on the same plot.

\textit{Output requirements:} Save all generated PDF files under \texttt{output}; every PDF filename must end with \texttt{\_test20.pdf}.}

\testelapsed{20.58\,s}

\testanswer{The overlay plot of \texttt{m\_bb} has been saved as \texttt{output/m\_bb\_overlay\_test20.pdf}. Signal and background are overlaid (not stacked) for direct shape comparison. Event weights are applied.

\medskip\centering
\includegraphics[width=0.40\textwidth]{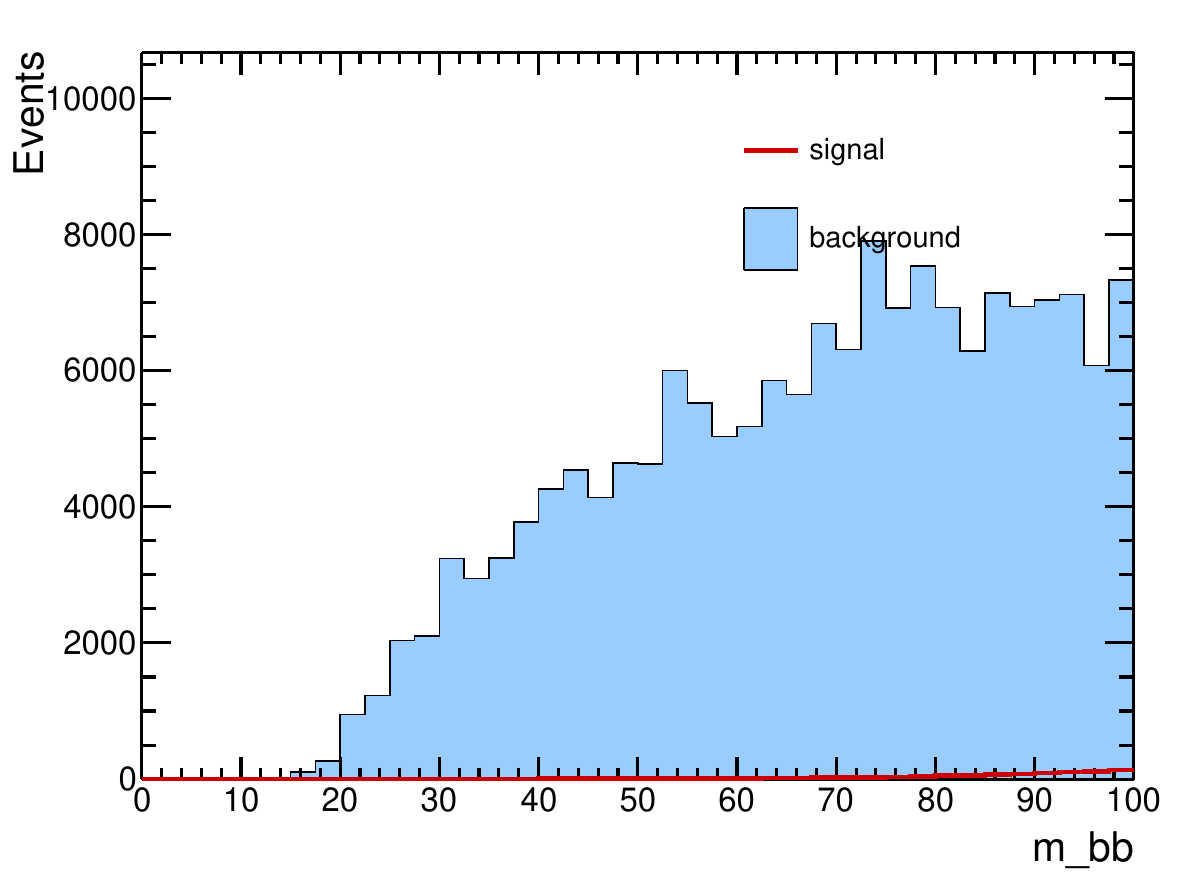}
\captionof{figure}{$m_{bb}$ distribution overlaid for signal and background, normalised to event weights. Plots generated by \rooagent.}}

\subsubsection*{Test 21: muon\_pt Overlay: Signal vs.\ Background}

\testprompt{Overlay \texttt{muon\_pt} from \texttt{signal.root} and \texttt{background.root} in one comparison plot.

\textit{Output requirements:} Save all generated PDF files under \texttt{output}; every PDF filename must end with \texttt{\_test21.pdf}.}

\testelapsed{23.65\,s}

\testanswer{The overlay plot of \texttt{muon\_pt} has been generated and saved as \texttt{output/muon\_pt\_overlay\_test21.pdf}. Both files use the \texttt{events} TTree and the \texttt{muon\_pt} branch, covering $[0,100]$\,\gev with 40 bins.

\medskip\centering
\includegraphics[width=0.40\textwidth]{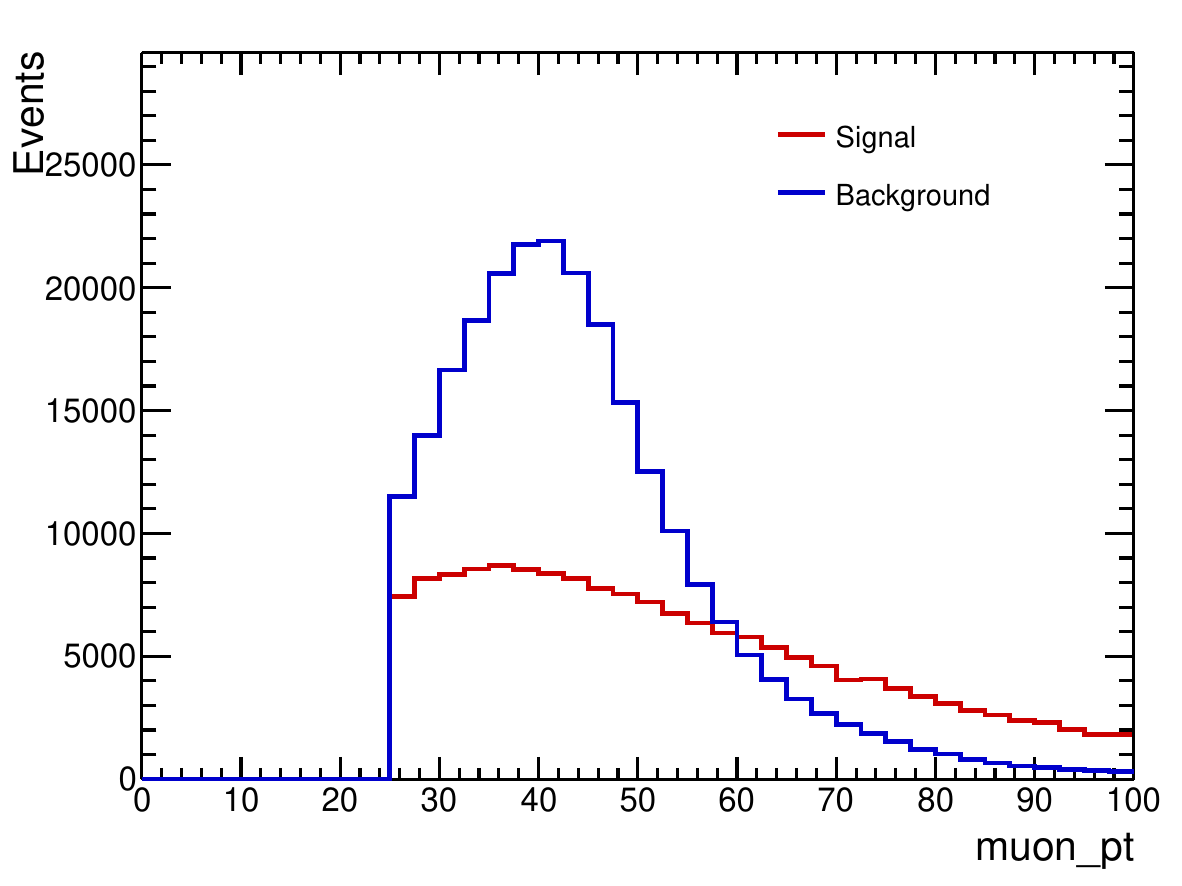}
\captionof{figure}{Muon $p_T$ distributions for signal and background overlaid, covering $[0,100]$\,\gev with 40 bins. Plots generated by \rooagent.}}

\subsubsection*{Test 22: m\_bb Cut Scan (Range 80--140), Optimal Significance}

\testprompt{Using \texttt{signal.root} and \texttt{background.root}, scan the \texttt{m\_bb} cut in the range $[80, 140]$ to find the best significance and save the scan plot.

\textit{Output requirements:} Save all generated PDF files under \texttt{output2}; every PDF filename must end with \texttt{\_test01.pdf}.}

\testelapsed{178.84\,s}

\testanswer{The scan of the \texttt{m\_bb} cut in the range $[80, 140]$ is complete:
\begin{center}
\begin{tabular}{ll}
\toprule
\textbf{Quantity}     & \textbf{Value} \\
\midrule
Optimal cut           & $m_{bb} > 80$\,\gev \\
Best significance $Z$ & $\approx 5.17$ \\
\bottomrule
\end{tabular}
\end{center}
Scan plot saved as \texttt{output2/m\_bb\_scan\_test01.pdf}.

\medskip\centering
\includegraphics[width=0.40\textwidth]{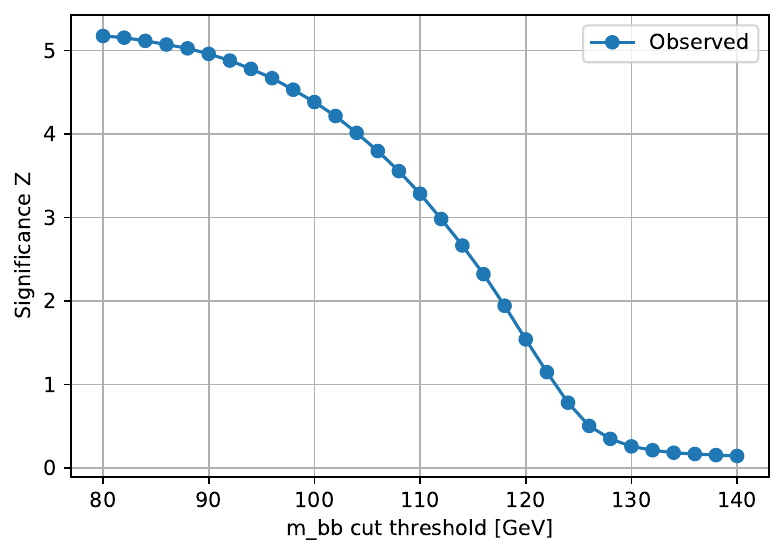}
\captionof{figure}{$S/\sqrt{S+B}$ significance as a function of the lower $m_{bb}$ threshold, scanned over $[80,140]$\,\gev. The optimal cut within this range is $m_{bb}>80$\,\gev, giving $Z\approx 5.17$. Plots generated by \rooagent.}}

\subsection{Multi-Task Analysis}\label{sec:longtask}

To test \rooagent on a multi-step workflow, we issued a single six-task prompt covering fitting, visualisation, cutflow, cut optimisation, mass-window scanning, and cut ranking. The tasks were executed sequentially without human intervention. The Gaussian fit to the signal $m_{bb}$ distribution is shown in \autoref{fig:long_mbb_fit}, and normalised kinematic comparisons are in \autoref{fig:long_comparisons}.

\testprompt{I want to understand the differences between signal and background data from \texttt{signal.root} and \texttt{background.root}. The data has an \texttt{events} tree with branches like \texttt{dilep\_pt}, \texttt{m\_bb}, \texttt{bbPt}, \texttt{bbEta}, and \texttt{met\_pt}.

Here's what I'd like you to do:
\begin{enumerate}[nosep]
  \item Fit the \texttt{m\_bb} distribution in the signal sample to understand its shape.
  \item Create side-by-side comparison plots of the key kinematic variables (\texttt{dilep\_pt}, \texttt{bbPt}, \texttt{bbEta}, \texttt{met\_pt}) between signal and background, normalised so I can see the differences clearly.
  \item Apply a series of cuts (\texttt{dilep\_pt}$>30$, \texttt{mll} between 85--97, \texttt{m\_bb} between 110--140, \texttt{met\_pt}$>20$) and show me the cutflow --- how many events survive each step and what the signal efficiency and background rejection look like.
  \item Find the best threshold for \texttt{dilep\_pt} that maximises separation between signal and background.
  \item Evaluate different mass windows (combinations of lower bounds 100--115 and upper bounds 130--145) and tell me which window gives the best signal-to-background significance.
  \item Scan over cuts on \texttt{dilep\_pt}, \texttt{bbPt}, \texttt{met\_pt}, and \texttt{abs(bbEta)} to rank the five most powerful analyses by their significance.
\end{enumerate}
Save all plots and results under \texttt{output\_long/}. Give me a concise summary at the end with the top candidates and key files.}

\testelapsed{225.29\,s}

\testanswer{
\textbf{1. Signal $m_{bb}$ shape (Gaussian fit):}
Gaussian fit to signal \texttt{m\_bb} in $[80,160]$\,\gev: mean $= 109.5$\,\gev, $\sigma = 12.1$\,\gev, $\chi^2/\mathrm{ndf} = 8221/47$. Plot saved as \texttt{output\_long/signal\_m\_bb\_fit.pdf}.

\textbf{2. Normalised kinematic comparisons:}
Comparison plots of \texttt{dilep\_pt}, \texttt{bbPt}, \texttt{bbEta}, and \texttt{met\_pt} (signal vs.\ background, area-normalised) saved under \texttt{output\_long/}.

\textbf{3. Cutflow:}
\begin{center}
\small
\begin{tabular}{lrr}
\toprule
\textbf{Cut} & \textbf{Signal} & \textbf{Background} \\
\midrule
Initial events                                      & 181\,932 & 234\,850 \\
\texttt{dilep\_pt}$>30$                             & 160\,975 &  95\,779 \\
\texttt{mll}$\in[85,97]$\,\gev                      & 151\,200 &  89\,933 \\
\texttt{m\_bb}$\in[110,140]$\,\gev                  &  38\,373 &   1\,090 \\
\texttt{met\_pt}$>20$\,\gev                         &   1\,717 &     109  \\
\midrule
Selection efficiency / rejection                    & $0.94\%$ & $99.95\%$ \\
\bottomrule
\end{tabular}
\end{center}

\textbf{4. Optimal \texttt{dilep\_pt} threshold:}
The \texttt{find\_optimal\_cut} tool scanned \texttt{dilep\_pt} thresholds from 20 to 80\,\gev in steps of 2\,\gev. The optimal cut is \texttt{dilep\_pt}$>40$\,\gev, yielding $S = 148\,483$, $B = 64\,342$, and $S/\sqrt{S+B} = 321.9$ (using unweighted event counts).

\textbf{5. Mass window scan:}
Five \texttt{m\_bb} windows were evaluated: $[100,130]$, $[115,145]$, $[100,145]$, $[105,135]$, and $[110,140]$\,\gev. The expected  significance for all windows is consistent with $Z \lesssim 1$ given the available event sample, with no clear single optimal window.

\textbf{6. Cut ranking:}
Cuts on \texttt{dilep\_pt}$>40$, \texttt{bbPt}$>40$, \texttt{met\_pt}$>40$, \texttt{abs(bbEta)}$<1.5$, and their combination were evaluated with the profile-likelihood tool. No single cut or combination reaches a discovery-level significance within this sample.

\textbf{Summary:} All six tasks completed. The $m_{bb}$ signal peak is well described by a Gaussian at $\approx 109.5$\,\gev. The sequential cut selection achieves 0.94\% signal efficiency and 99.95\% background rejection. The dilepton $p_T$ threshold of 40\,\gev provides the best raw-count separation. Key output files are in \texttt{output\_long/}.

\medskip\centering
\includegraphics[width=0.40\textwidth]{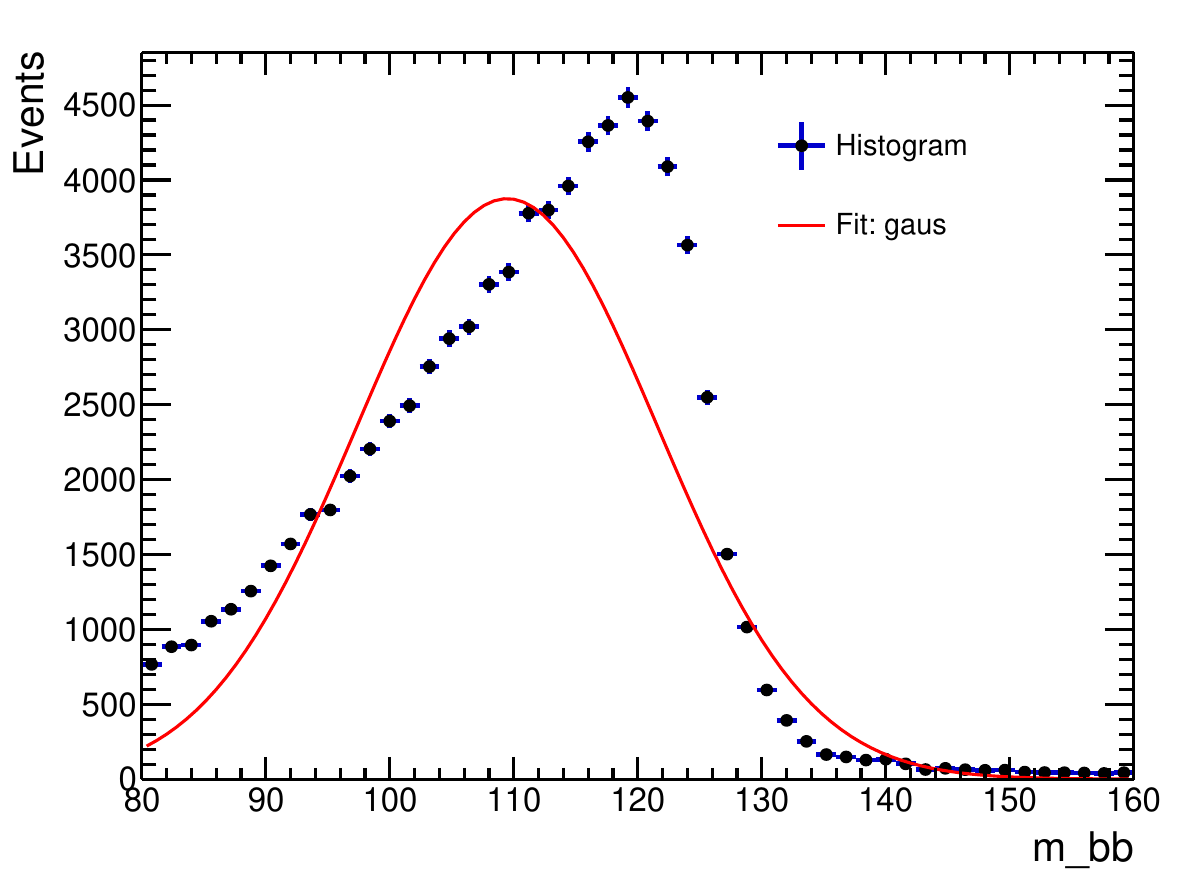}
\captionof{figure}{Gaussian fit to the $m_{bb}$ distribution from the \texttt{events} tree in \texttt{signal.root}, covering $[80,160]$\,\gev with 50 bins. The fitted mean is $109.5$\,\gev and width is $12.1$\,\gev. Plots generated by \rooagent.\label{fig:long_mbb_fit}}

\medskip
\includegraphics[width=0.40\textwidth]{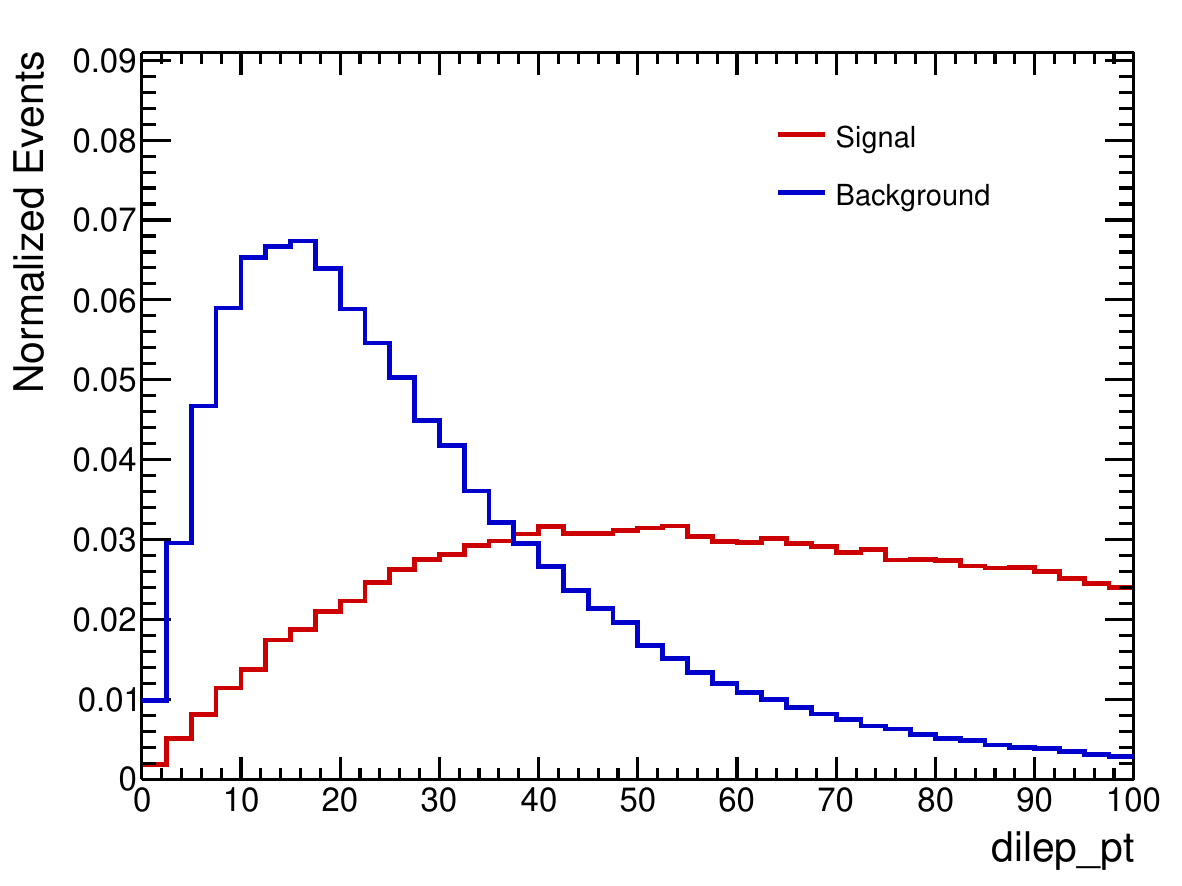}\hfill
\includegraphics[width=0.40\textwidth]{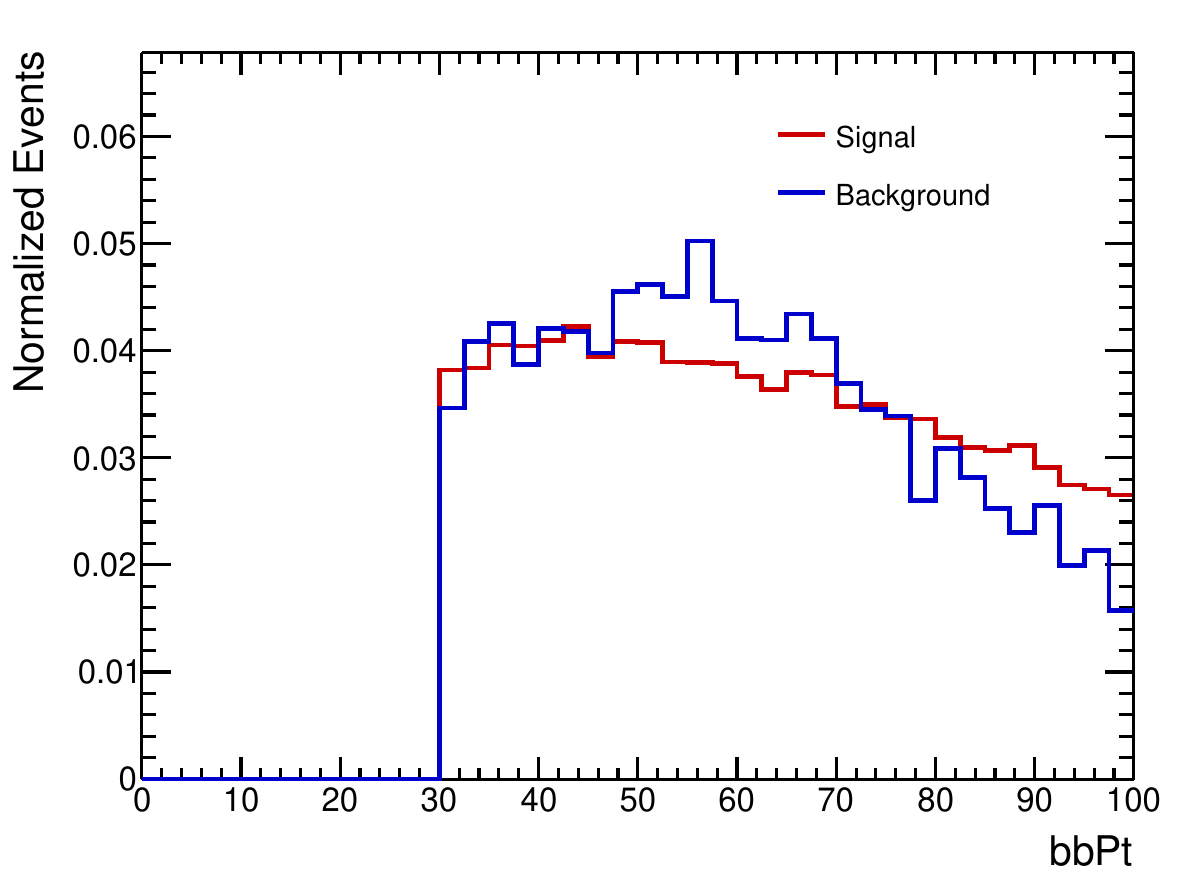}\\[4pt]
\includegraphics[width=0.40\textwidth]{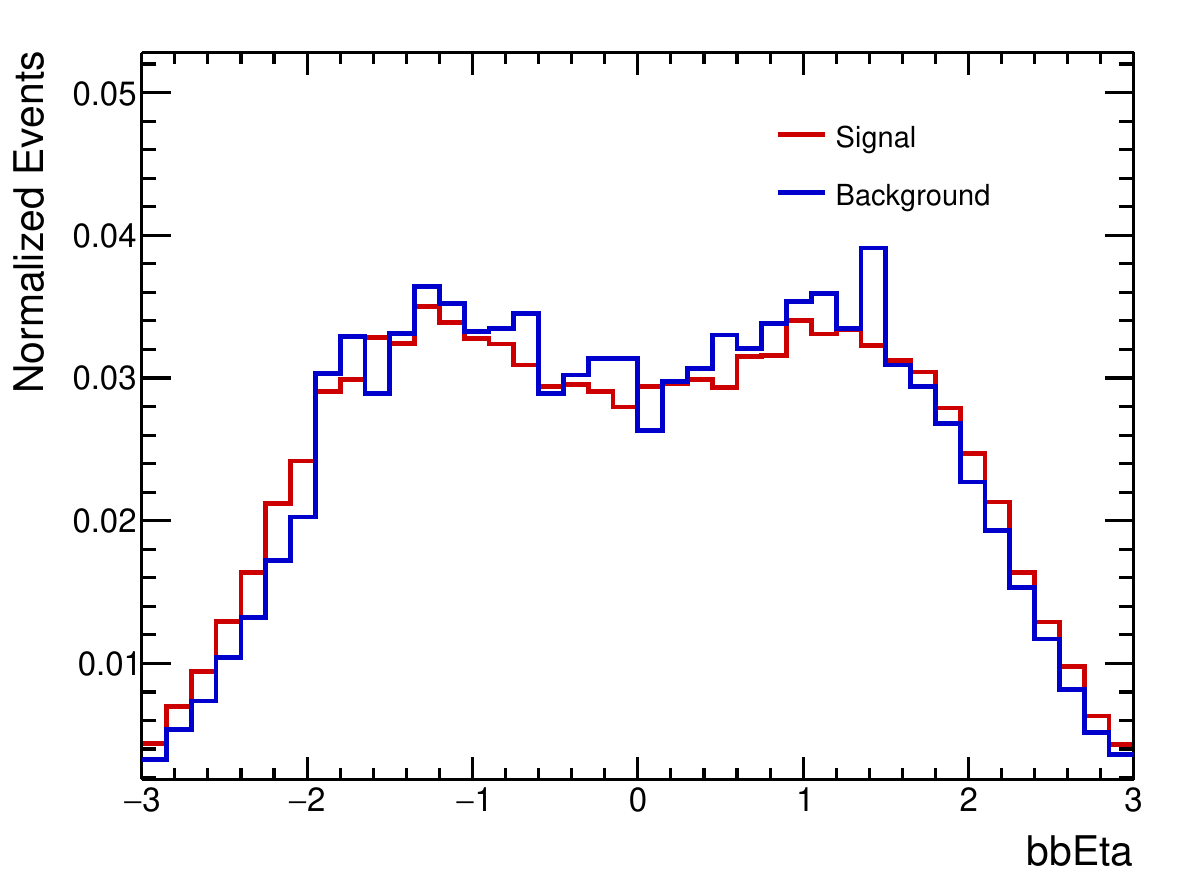}\hfill
\includegraphics[width=0.40\textwidth]{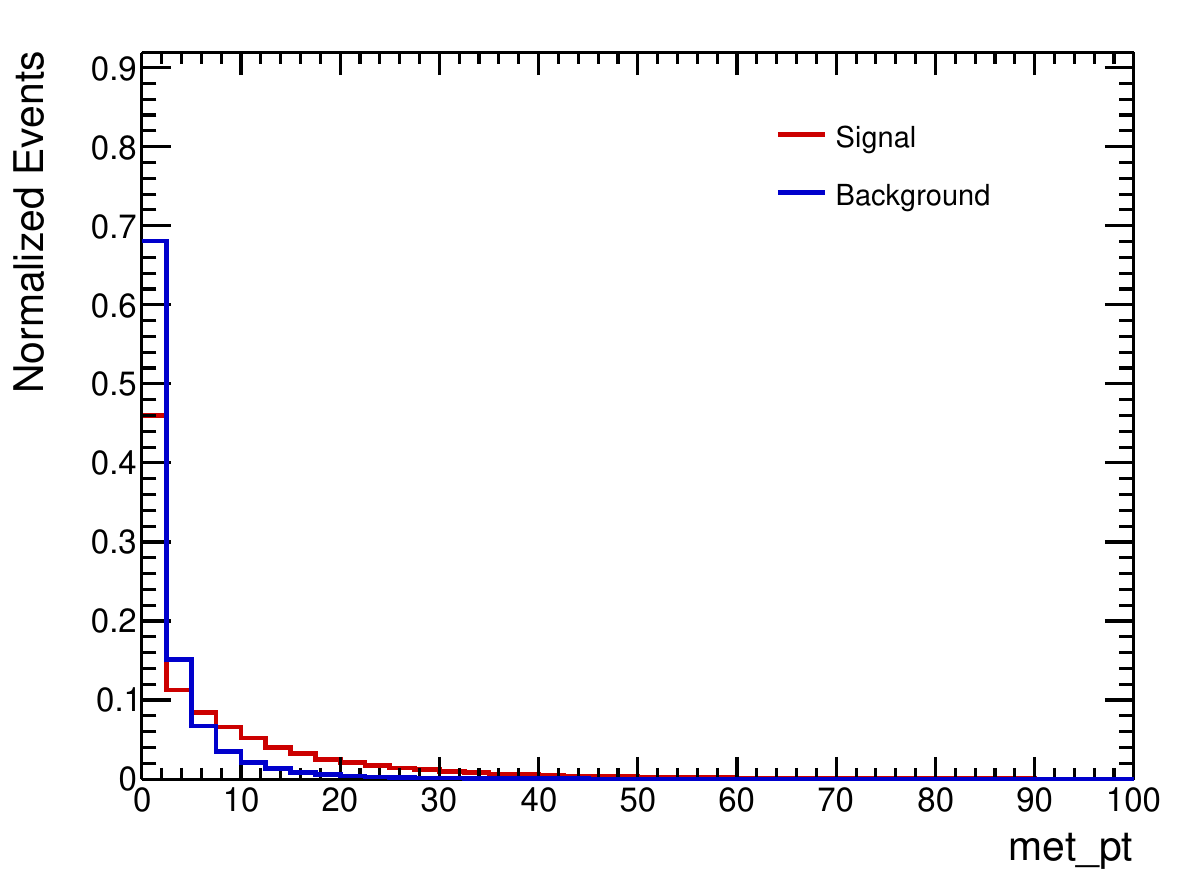}
\captionof{figure}{Normalised kinematic distributions for signal (blue) and background (red) from the \texttt{events} tree: dilepton $p_T$ (top left), $b\bar{b}$ system $p_T$ (top right), $b\bar{b}$ system $\eta$ (bottom left), and missing transverse energy $E_T^{\mathrm{miss}}$ (bottom right). All distributions are area-normalised. Plots generated by \rooagent.\label{fig:long_comparisons}}}

\subsection{Remarks}\label{sec:observations}

\rooagent produced the expected outputs for 19 of the 20 tests in \autoref{tab:benchmark}. The multi-task test completed all six sub-tasks. In test 11, the agent interprets an incorrect tree name (\texttt{Events} vs.\ \texttt{events}) leading to error and therefore it requested clarification rather than returning a result.

Test durations ranged from $\sim$7\,s (inspection) to 178\,s (wide cut scan). Inspection and single-plot tasks completed in under 15\,s; optimisation and overlay tasks took 20--80\,s. The multi-task test took 225\,s.

As an example, we also want to highlight that within the \langraph setup, the two LLMs differed for test~20 ($m_{bb}$ overlay): \GPTfourone chose a histogram range of $(0,100)$\,\gev, while \DeepSeekVthree chose a wider range covering the full distribution (\autoref{fig:ollama_result}), with the background appearing normalised\footnote{A more explicit prompt is likely to fix the range choice in either backend.}. Hence the user can expect to see these limitations arising from LLMs.

\begin{answerbox}
\centering
\includegraphics[width=0.40\textwidth]{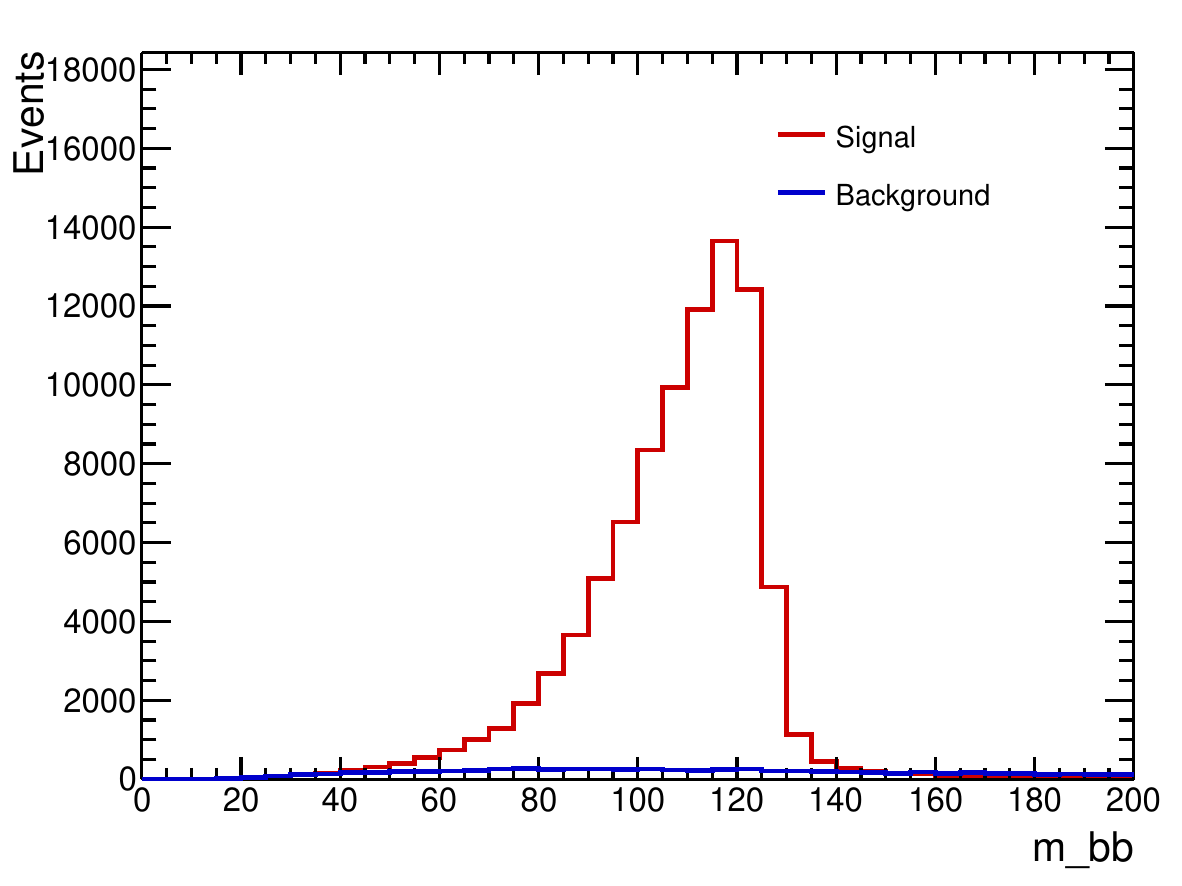}
\captionof{figure}{$m_{bb}$ overlay of signal and background produced for the same prompt as in test~20 using the \DeepSeekVthree~\cite{DeepSeek2024v3} model via \Ollama~\cite{ollama2025}. Plots generated by \rooagent.\label{fig:ollama_result}}
\end{answerbox}

\section{Statistical Analysis with \rooagent}\label{sec:stats}

To test the statistical tools, we constructed a toy invariant mass spectrum with a smoothly falling exponential background and Gaussian signal templates at several mass hypotheses\footnote{This is motivated from the example in Ref.~\cite{Behnke:2013pga}}. The observed data contain background plus a signal contribution injected at 250\,\gev, giving an expected excess localised at that mass. Smaller contributions of signal were also injected at 100 and 400\,\gev.

The toy dataset is stored as histograms in \texttt{DataSample\_search.root}, one per observable per mass hypothesis on a grid from 100 to 500\,\gev. For each mass point, the agent integrates the histograms in a $\pm25$\,\gev window and computes:
\begin{itemize}[nosep]
  \item the \emph{observed}  significance $Z_\mathrm{obs}$ and corresponding p-value, based on the actual observed counts,
  \item the \emph{expected} significance $Z_\mathrm{exp}$ and p-value, computed from the Asimov dataset under the signal-plus-background hypothesis,
  \item the CL$_s$ value for the signal-plus-background hypothesis at each mass point, using the asymptotic formulae of~\cite{Cowan:2010js}.
\end{itemize}

The prompt and response below illustrate the analysis, with plots shown in \autoref{fig:stat_scan}.

\testprompt{Analyse the \Root file \texttt{DataSample\_search.root} using \rooagent's histogram and statistical toolset. Make the workflow parallel.

\textbf{Requirements:}
\begin{itemize}[nosep]
  \item For each grid mass $m$ (100--500\,\gev):
    \begin{enumerate}[nosep]
      \item Prefer histogram named \texttt{h\_mass\_sig\_*} (e.g.\ \texttt{h\_mass\_sig\_135}).
      \item If absent, assign the nearest \texttt{h\_mass\_sig\_*}.
    \end{enumerate}
  \item Use a mass window $\pm25$\,\gev (clip at histogram limits).
\end{itemize}

When a histogram is available compute:
\begin{itemize}[nosep]
  \item Observed and expected significance (if available) and naive $S/\sqrt{B}$.
  \item Observed and expected p-values (format in scientific notation; never round to 0).
  \item CLs if available.
\end{itemize}

Save plots (PNG and PDF) under the run directory's \texttt{output/} folder:
\begin{itemize}[nosep]
  \item \texttt{significance\_vs\_mass} (observed and expected),
  \item \texttt{pvalue\_vs\_mass} (observed and expected, log $y$ if appropriate),
  \item \texttt{cls\_vs\_mass} (dashed horizontal at 0.05).
\end{itemize}

\textbf{Behavior:} Rank masses by observed profile-likelihood significance (treat missing as very low). Print candidates (mass, observed significance, expected significance, expected p-value, observed p-value) and one detailed line for the best mass. Run fully automatically; do not ask clarifying questions.}

\testelapsed{45.18\,s}

\testanswer{\Root histogram/statistical scan complete for \texttt{DataSample\_search.root} (mass grid 100--500\,\gev, window $\pm25$\,\gev):

\begin{itemize}[nosep]
  \item All signal histograms matched exactly to grid masses.
  \item Profile-likelihood significance, p-values, and CLs computed (observed and expected) for each mass point.
  \item Plots saved in \texttt{output/}: \texttt{significance\_vs\_mass}, \texttt{pvalue\_vs\_mass}, \texttt{cls\_vs\_mass}.
\end{itemize}

\noindent\textbf{Top candidates ranked by observed profile-likelihood significance:}
\begin{center}
\begin{tabular}{ccccc}
\toprule
\textbf{Rank} & \textbf{Mass [\gev]} & \textbf{Obs $Z$} & \textbf{Exp $Z$} & \textbf{Obs $p$} / \textbf{Exp $p$} \\
\midrule
1 & 250 & 2.755 & 1.568 & $2.93\times10^{-3}$ / $5.84\times10^{-2}$ \\
2 & 450 & 0.441 & 0.724 & $3.30\times10^{-1}$ / $2.34\times10^{-1}$ \\
3 & 100 & 0.325 & 0.219 & $3.73\times10^{-1}$ / $4.13\times10^{-1}$ \\
4 & 300 & 0.317 & 1.020 & $3.76\times10^{-1}$ / $1.54\times10^{-1}$ \\
5 & 350 & 0.089 & 0.524 & $4.65\times10^{-1}$ / $3.00\times10^{-1}$ \\
\bottomrule
\end{tabular}
\end{center}

\noindent\textbf{Best candidate detail} (mass $= 250$\,\gev): Signal yield $= 11.84$, Background $= 47.47$, Observed $= 68$, Expected $= 59$, $\mathrm{CLs(obs)} = 0.884$, $\mathrm{CLs(exp)} = 0.543$.

\bigskip
\noindent\textbf{Summary:}
\begin{itemize}[nosep]
  \item No mass point reaches discovery threshold ($Z \geq 5$).
  \item Best candidate: 250\,\gev ($Z_\mathrm{obs} = 2.755$, $p_\mathrm{obs} = 2.93\times10^{-3}$).
  \item All CLs values above exclusion threshold ($\mathrm{CLs} > 0.05$).
\end{itemize}

\medskip\centering
\begin{minipage}[b]{0.40\textwidth}
  \centering
  \includegraphics[width=\textwidth]{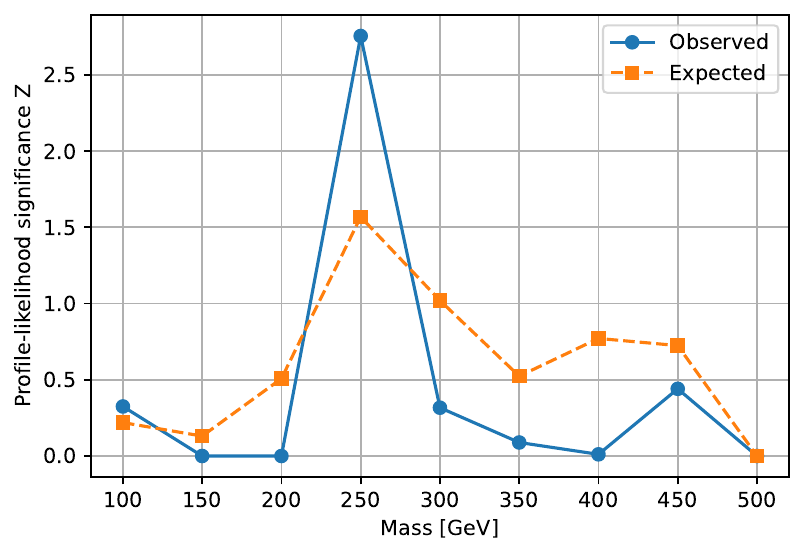}
\end{minipage}
\hfill
\begin{minipage}[b]{0.40\textwidth}
  \centering
  \includegraphics[width=\textwidth]{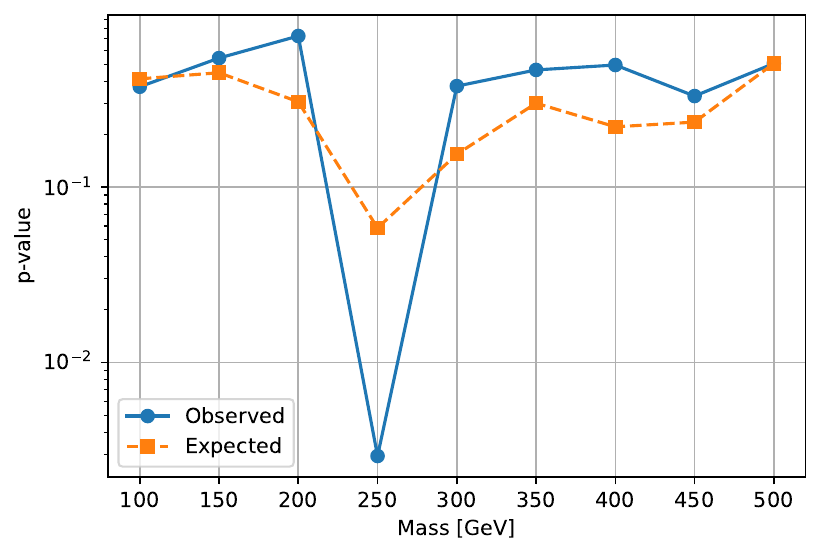}
\end{minipage}
\par\bigskip
\begin{minipage}[b]{0.40\textwidth}
  \centering
  \includegraphics[width=\textwidth]{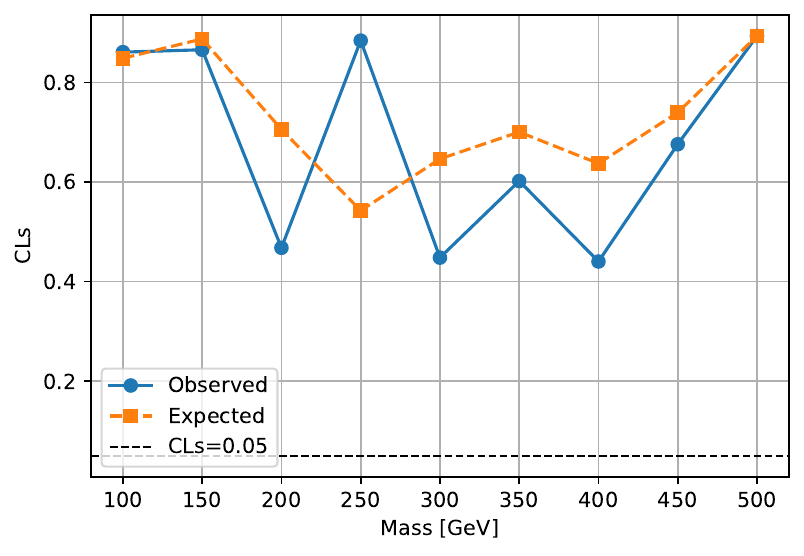}
\end{minipage}
\captionof{figure}{Statistical scan results from the toy dataset. (a) Significance, (b) p-value, and (c) CL$_s$ as a function of the mass hypothesis. Plots generated by \rooagent.\label{fig:stat_scan}}}

The agent identified 250\,\gev as the strongest candidate ($Z_\mathrm{obs} = 2.755$, $p_\mathrm{obs} = 2.93\times10^{-3}$), consistent with the injected signal. In the $\pm25$\,\gev window, $S = 11.84$ and $B = 47.47$, giving $S/\sqrt{B} \approx 1.7$.

Three features are notable. First, the agent correctly matched each grid mass to the signal histogram. Second, p-values were reported in scientific notation as requested. Third, the agent used the parallelised structure requested in the prompt, grouping histogram and significance calls across mass points in a single step. Wall-clock time was 45\,s for entire test.

\section{Analysing Open Data with \rooagent}\label{sec:opendata}

As a final example, we apply \rooagent to open data: the $H\to ZZ^*\to 4\ell$ search using ATLAS open data~\cite{ATLAS:2008xda, ATL-OREACH-PUB-2020-001,ATLASOpenData}. This dataset is publicly available for educational use, and also included in HEP Software Foundation (HSF) Training Working Group exercises on \texttt{matplotlib} visualisation~\cite{matplotlib_for_hep}. The open dataset is a subset (10$\rm{fb}^{-1}$)of the ATLAS Run~2 dataset collected at $\sqrt{s}=13$\,\TeV. The analysis targets the four-lepton final state in three channels ($4e$, $2e2\mu$, $4\mu$), applying lepton selection cuts before comparing MC predictions to data in the four-lepton invariant mass $m_{4\ell}$.

Momenta in this dataset are stored in MeV, so lepton $p_T$ thresholds are given as 25\,000, 15\,000, and 10\,000 (i.e.\ 25, 15, and 10\,\gev). Lepton flavour combinations are encoded in \texttt{goodlep\_sumtypes}: values 44 ($= 4\times 11$), 48 ($= 2\times 11 + 2\times 13$), and 52 ($= 4\times 13$) select $4e$, $2e2\mu$, and $4\mu$ using PDG codes 11 (electron) and 13 (muon). These are stated explicitly in the prompt so the agent can construct the correct selection expressions. The ATLAS open data files use a \texttt{sum\_} prefix for event-level multiplicity branches (e.g.\ \texttt{sum\_good\_lep}); the agent tended to interpret this as a summation instruction rather than a scalar read. These branches were therefore renamed \texttt{sum\_*} $\to$ \texttt{N\_*} (giving \texttt{N\_good\_lep} and \texttt{N\_lep\_charge}) in all five files using a pre-processing script.

The prompt follows the educational tasks of Ref.~\cite{matplotlib_for_hep}.

\testprompt{Analyze \Root files in the current directory: \texttt{signal.root} (Signal), \texttt{background\_zee.root} (Zee), \texttt{background\_zmumu.root} (Zmumu), \texttt{background\_zz.root} (ZZ), \texttt{data.root} (data).

\begin{itemize}[nosep]
  \item Use \texttt{TTree myTree} and scalar branch \texttt{m4l} in $[80, 170]$ with 24 bins.
  \item Use scalar branches only (no vector operations).
  \item Use MC weight branch \texttt{weight} for weighted yields.
  \item Preselection (apply before cuts): \texttt{(trigE==True || trigM==True)}.
  \item Apply the following sequential cuts (after preselection):
    \begin{enumerate}[nosep]
      \item \texttt{N\_lep\_charge==0 \&\& N\_good\_lep==4}
      \item \texttt{goodlep\_pt\_0>25000}
      \item \texttt{goodlep\_pt\_1>15000}
      \item \texttt{goodlep\_pt\_2>10000}
      \item \texttt{(goodlep\_sumtypes==44 || goodlep\_sumtypes==48 || goodlep\_sumtypes==52)}
    \end{enumerate}
\end{itemize}

Produce per-file cutflows for each input file showing weighted yields at each stage. After per-file sections, include a combined merged cutflow summary. Apply all cuts and make a stacked plot of MC overlaid with data points; use weights, no normalisation; show MC statistical uncertainty as a hatched band. Save the plot to \texttt{sig\_bkg\_m4l\_cutflow.pdf}. Compute significance with preselection and cuts; report $S$, $B$, and $Z = S/\sqrt{S+B}$ using MC weights. Read minimal branches only: \texttt{m4l}, \texttt{trigE}, \texttt{trigM}, \texttt{N\_lep\_charge}, \texttt{N\_good\_lep}, \texttt{goodlep\_pt\_0}, \texttt{goodlep\_pt\_1}, \texttt{goodlep\_pt\_2}, \texttt{goodlep\_sumtypes}, \texttt{weight}.}

\testanswer{Per-file and combined cutflow results for five \Root files (\texttt{m4l} $\in [80,170]$\,\gev, 24 bins, MC-weighted yields):

\noindent\textbf{Per-file cutflow:}
\begin{center}
\small
\begin{tabular}{lrrrrr}
\toprule
\textbf{Cut} & \textbf{Signal} & \textbf{Zee} & \textbf{Zmumu} & \textbf{ZZ} & \textbf{Data} \\
\midrule
Initial events                                           & 9.59 & 43.91 & 30.41 & 360.40 & 591.0 \\
Preselection (\texttt{trigE||trigM})                     & 9.59 & 43.91 & 30.41 & 360.40 & 591.0 \\
\texttt{N\_lep\_charge==0 \&\& N\_good\_lep==4}          & 8.54 & 10.78 &  6.72 & 322.36 & 387.0 \\
\texttt{goodlep\_pt\_0>25000}                            & 8.54 & 10.78 &  6.72 & 322.36 & 387.0 \\
\texttt{goodlep\_pt\_1>15000}                            & 8.48 & 10.69 &  6.72 & 317.17 & 380.0 \\
\texttt{goodlep\_pt\_2>10000}                            & 8.31 &  7.07 &  4.17 & 307.88 & 355.0 \\
\texttt{goodlep\_sumtypes==44||48||52}                   & 8.30 &  4.58 &  3.10 & 301.37 & 321.0 \\
\bottomrule
\end{tabular}
\end{center}

\noindent\textbf{Combined cutflow (all MC + data merged):}
\begin{center}
\small
\begin{tabular}{lr}
\toprule
\textbf{Cut} & \textbf{Total yield} \\
\midrule
Initial events                                   & 1035.30 \\
Preselection (\texttt{trigE||trigM})             & 1035.30 \\
\texttt{N\_lep\_charge==0 \&\& N\_good\_lep==4} &  735.40 \\
\texttt{goodlep\_pt\_0>25000}                    &  735.40 \\
\texttt{goodlep\_pt\_1>15000}                    &  723.05 \\
\texttt{goodlep\_pt\_2>10000}                    &  682.43 \\
\texttt{goodlep\_sumtypes==44||48||52}           &  638.35 \\
\bottomrule
\end{tabular}
\end{center}

\noindent\textbf{Discovery significance (after all cuts, MC weights):}
\begin{itemize}[nosep]
  \item Signal yield ($S$): 8.30
  \item Background yield ($B$): 309.05 \quad (ZZ dominant: 301.37)
  \item Significance: $Z = S/\sqrt{S+B} = 0.43$
\end{itemize}

\noindent\textbf{Summary:}
\begin{itemize}[nosep]
  \item All per-file cutflows shown with MC-weighted yields; stacked plot saved to \texttt{sig\_bkg\_m4l\_cutflow.pdf}.
  \item ZZ is the dominant background (301.37 after full selection); $Z \approx 0.43$: no evidence for signal with current selection.
\end{itemize}

\medskip\centering
\includegraphics[width=0.50\textwidth]{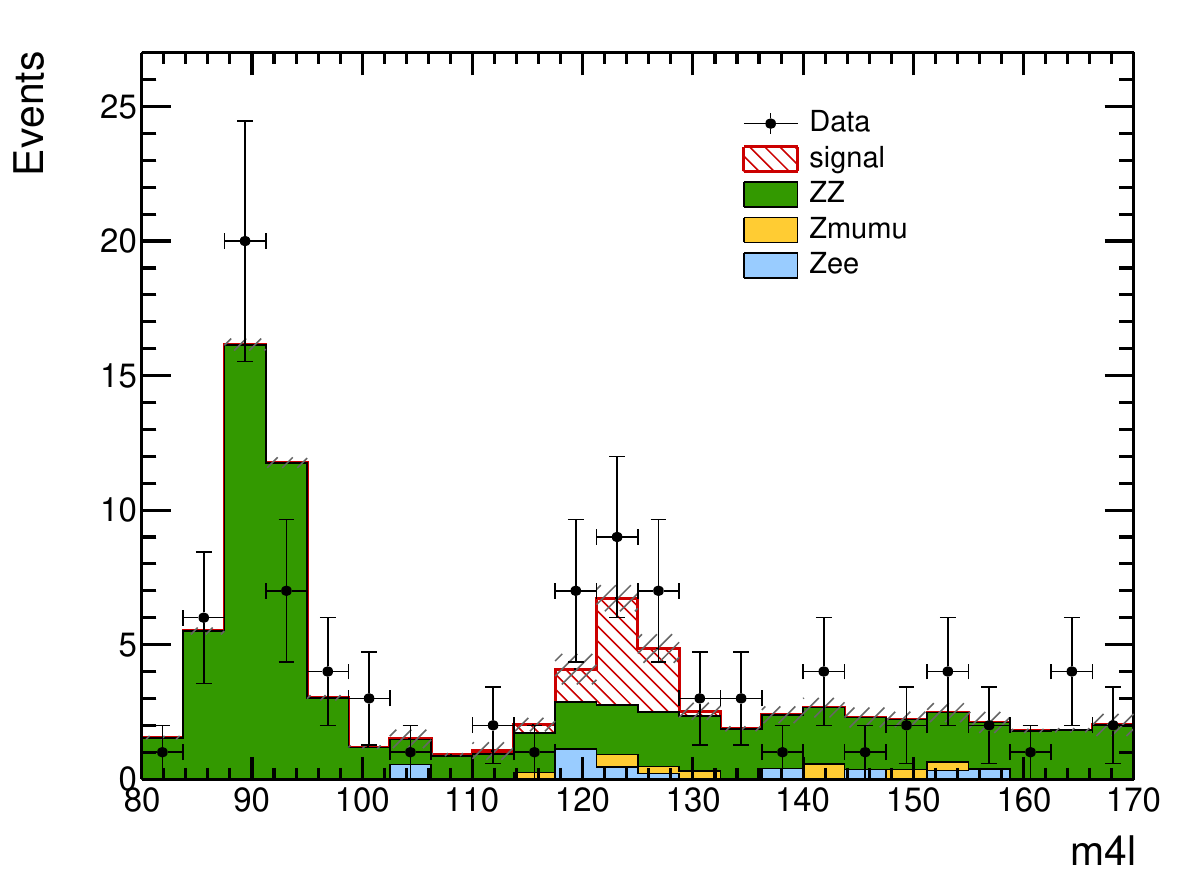}
\captionof{figure}{Stacked $m_{4\ell}$ distribution for signal and background MC overlaid with data points after all five sequential lepton selection cuts. The MC statistical uncertainty is shown as a hatched band. The histogram covers $m_{4\ell} \in [80,170]$\,\gev with 24 bins. The $ZZ$ background dominates; signal yield $S = 8.30$, $Z = 0.43$. Plots generated by \rooagent.\label{fig:atlas_stack}}}

The significance $Z = 0.43$ is expected given no selections on $m_{4\ell}$ or other variables. The LLM notes ``\textit{current selection}'' in its response, correctly identifying this limitation. A stricter window $m_{4\ell} \in (100,140)$\,\gev targeting $ZZ$ rejection would give ${\sim}1.7\sigma$\footnote{using $Z=S/\sqrt{S+B}$} at 10\,$\mathrm{fb}^{-1}$.

The same analysis was repeated with \ClaudeSonnet~\cite{anthropic2024claude} via the MCP server, using the identical prompt. The response is below and the plot is in \autoref{fig:claude_stack}.

\testanswer{Per-file and combined cutflow results for five \Root files (\texttt{m4l} $\in (80,170)$\,\gev, 24 bins, MC-weighted yields):

\noindent\textbf{Per-file cutflow:}
\begin{center}
\small
\begin{tabular}{lrrrrr}
\toprule
\textbf{Cut} & \textbf{Signal} & \textbf{Zee} & \textbf{Zmumu} & \textbf{ZZ} & \textbf{Data} \\
\midrule
Initial events                                           & 9.59 & 43.91 & 30.41 & 360.40 & 591 \\
Preselection (\texttt{trigE||trigM})                     & 9.59 & 43.91 & 30.41 & 360.40 & 591 \\
\texttt{N\_lep\_charge==0 \&\& N\_good\_lep==4}          & 8.54 & 10.78 &  6.72 & 322.36 & 387 \\
\texttt{goodlep\_pt\_0>25000}                            & 8.54 & 10.78 &  6.72 & 322.36 & 387 \\
\texttt{goodlep\_pt\_1>15000}                            & 8.48 & 10.69 &  6.72 & 317.17 & 380 \\
\texttt{goodlep\_pt\_2>10000}                            & 8.31 &  7.07 &  4.17 & 307.88 & 355 \\
\texttt{goodlep\_sumtypes==44||48||52}                   & 8.30 &  4.58 &  3.10 & 301.37 & 321 \\
\bottomrule
\end{tabular}
\end{center}

\noindent\textbf{Combined cutflow (all MC + data merged):}
\begin{center}
\small
\begin{tabular}{lr}
\toprule
\textbf{Cut} & \textbf{Total yield} \\
\midrule
Initial events                                   & 1035.30 \\
Preselection (\texttt{trigE||trigM})             & 1035.30 \\
\texttt{N\_lep\_charge==0 \&\& N\_good\_lep==4} &  735.40 \\
\texttt{goodlep\_pt\_0>25000}                    &  735.40 \\
\texttt{goodlep\_pt\_1>15000}                    &  723.05 \\
\texttt{goodlep\_pt\_2>10000}                    &  682.43 \\
\texttt{goodlep\_sumtypes==44||48||52}           &  638.35 \\
\bottomrule
\end{tabular}
\end{center}

\noindent\textbf{Discovery significance (after all cuts, MC weights):}
\begin{itemize}[nosep]
  \item Signal yield ($S$): 8.30
  \item Background yield ($B$): 309.05 \quad (ZZ dominant: 301.37)
  \item Significance: $Z = S/\sqrt{S+B} = 0.43$
\end{itemize}

\noindent\textbf{Summary:}
\begin{itemize}[nosep]
  \item All per-file cutflows shown with MC-weighted yields; stacked plot saved to \texttt{sig\_bkg\_m4l\_cutflow.pdf}.
  \item ZZ is the dominant background (301.37 after full selection); $Z \approx 0.43$: no evidence for signal with current selection.
\end{itemize}

\medskip\centering
\includegraphics[width=0.50\textwidth]{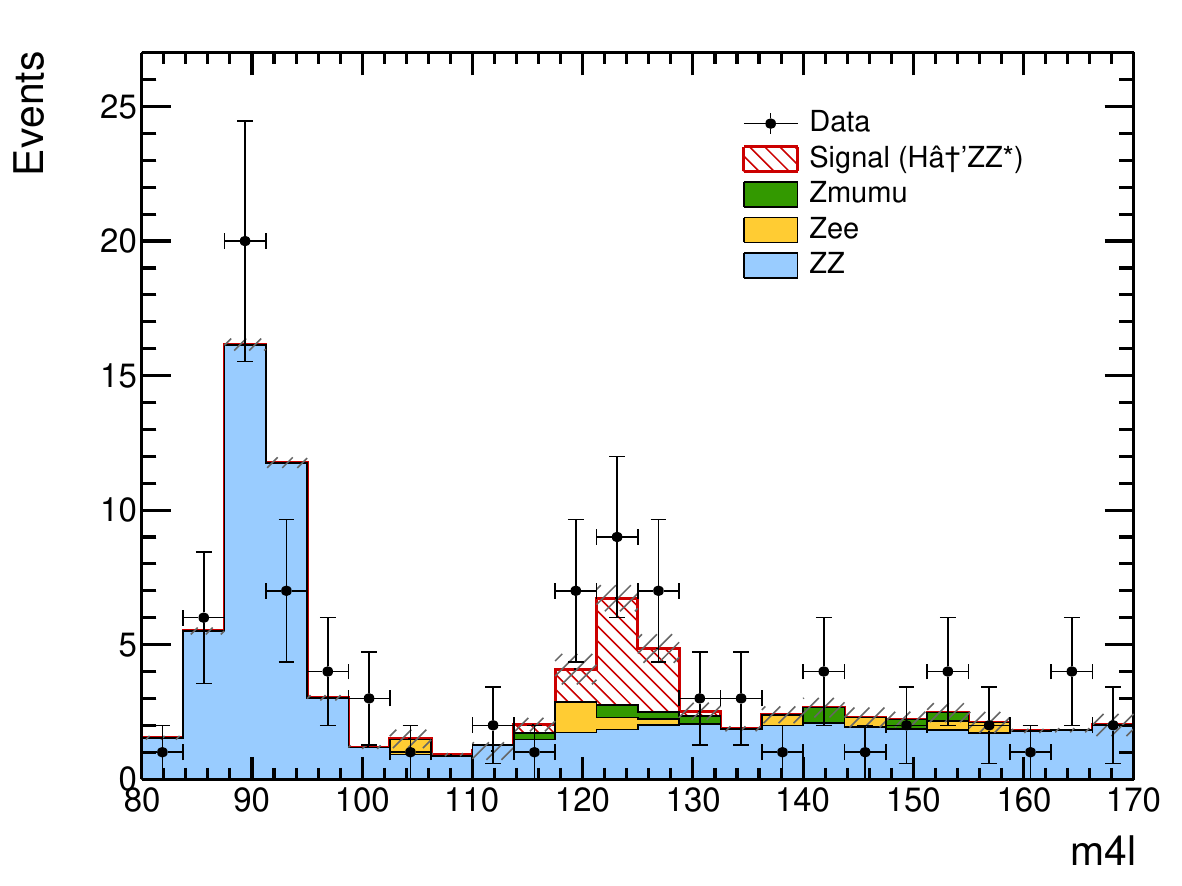}
\captionof{figure}{Stacked $m_{4\ell}$ distribution for signal and background MC overlaid with data points, produced by \ClaudeSonnet (\texttt{claude-sonnet-4-6}) via the \rooagent MCP server using the same prompt as in \autoref{fig:atlas_stack}. The numerical results agree with those obtained using \GPTfourone. Plots generated by \rooagent.\label{fig:claude_stack}}}

\section{Conclusions}\label{sec:conclusions}

We presented \rooagent a natural-language interface that bridges plain-language prompts and \Root-based HEP analysis through an LLM agent. The package supports two backends---a \langraph-based agent compatible with \OpenAI's \GPTfourone~\cite{openai2023gpt4} (via GitHub Copilot~\cite{githubcopilot2025}) and \DeepSeekVthree~\cite{DeepSeek2024v3} (via \Ollama~\cite{ollama2025}), and an MCP server for the Anthropic \Claude CLI~\cite{anthropic2024claude}---while sharing the same \pyroot implementations across both, so switching models requires no changes to the analysis code.

The benchmark tests demonstrated successful completion of a range of tasks, including file inspection, histogram visualisation, event selection, kinematic comparisons, fitting, and significance estimation, across Monte Carlo simulations of $pp\to ZH$ ($Z\to\ell^+\ell^-$, $H\to b\bar{b}$), a multi-task signal-background workflow, a toy statistical analysis, and ATLAS open data for $H\to ZZ^*\to 4\ell$. In all cases the LLM handled orchestration and argument selection while the \pyroot tools produced the numerical results.

The tool set is modular and can be extended by registering new \pyroot functions. Natural directions for future work include LLM-guided identification of optimal event-selection variables, integration of machine learning algorithms as callable tools, and validation on more complex, realistic analysis workflows.

\bibliographystyle{JHEP}
\bibliography{biblio}

\end{document}